\documentclass[pdftex,twocolumn,epjc3]{svjour3}          

\RequirePackage[T1]{fontenc}

\RequirePackage{graphicx}
\RequirePackage{mathptmx}      
\RequirePackage{flushend}
\RequirePackage[numbers,sort&compress]{natbib}
\RequirePackage[colorlinks,citecolor=blue,urlcolor=blue,linkcolor=blue]{hyperref}
\usepackage{ulem}

\usepackage{cuted}
\usepackage{xcolor}
\usepackage{amsmath}
\usepackage{caption}
\usepackage{subcaption}
\usepackage{verbatim}
\usepackage{xcolor}
\usepackage{orcidlink}
\usepackage{booktabs}

\renewcommand{\eqref}[1]{Eq.~(\ref{#1})}
\renewcommand{\Re}{\mathrm{Re}}

\journalname{Eur. Phys. J. C}

\begin{document}

\title{Z' boson mass reach and model discrimination at muon colliders}

\author{Kateryna Korshynska\thanksref{e1,addr1, addr3, addr4}
        \,\orcidlink{0000-0001-7469-746X}
\and
        Maximilian L\"oschner\thanksref{e2,addr1}
        \,\orcidlink{0000-0002-0850-257X}
        \and
        Mariia Marinichenko\thanksref{e3,addr1,addr3,addr5}
        \,\orcidlink{0009-0003-8861-636X}
        \and \\
        Krzysztof M\k{e}ka{\l}a\thanksref{e4,addr1,addr2}
        \,\orcidlink{0000-0003-4268-508X}
        \and
        J\"urgen Reuter\thanksref{e5,addr1}
        \,\orcidlink{0000-0003-1866-0157}
}

\thankstext{e1}{\textit{kateryna.korshynska@desy.de}}
\thankstext{e2}{\textit{maximilian.loeschner@desy.de}}
\thankstext{e3}{\textit{mariia.marinichenko@desy.de}}
\thankstext{e4}{\textit{krzysztof.mekala@desy.de}}
\thankstext{e5}{\textit{juergen.reuter@desy.de}}

\institute{Deutsches Elektronen-Synchrotron DESY, Notkestr. 85, 22607 Hamburg, Germany\label{addr1}
          \and
          Department of Physics, Taras Shevchenko National University of Kyiv, 
        64/13, Volodymyrska Street, 01601 Kyiv, Ukraine \label{addr3}
        \and
        Fundamentale Physik für Metrologie FPM, Physikalisch-Technische Bundesanstalt PTB, Bundesallee 100, 38116 Braunschweig, Germany \label{addr4}
        \and
        Instituut-Lorentz, Leiden University, Niels Bohrweg 2, 2333 CA Leiden, The Netherlands \label{addr5}
        \and
          Faculty of Physics, University of Warsaw, Pasteura 5, 02-093 Warszawa, Poland\label{addr2}
}

\date{Received: 4 March 2024 / Accepted: 8 May 2024}

\maketitle

\begin{abstract}
We study the discrimination power of future multi-TeV muon colliders for a large set of models with extended gauge symmetries and additional neutral gauge bosons (``$Z'$-models'').
Our study is carried out using a $\chi^2$-analysis of leptonic observables of s-channel scattering in effective $Z'$-models.
We make use of angular and chiral asymmetries induced in such models to find the discovery reach of a given muon collider setup in terms of the $Z'$ mass and to discriminate between the different scenarios.
In this context, we discuss how polarized beams --  should they become available at muon colliders -- or polarization measurements can help in the discrimination.
Our results show that typical muon collider setups which are currently under consideration can give a significantly higher reach compared to existing bounds and projections for high-luminosity LHC.
\end{abstract}

\sloppy

\section{Motivation}
The Standard Model (SM) of particle physics is among the most well-tested theories to date and can explain experimental observations to unprecedented precision despite thorough searches for deviations at modern collider experiments.
Nonetheless, several well-known phenomena such as the nature of dark matter, the baryon-antibaryon asymmetry of the universe or even the universal law of gravity lack an explanation within this theory.
To accommodate for effects Beyond the Standard Model (BSM), multiple extensions of the current formulation have been proposed.
If such extensions are based on new gauge symmetries, the existence of additional neutral gauge bosons (which we denote as $Z'$) is the natural consequence.
In the absence of a unique well-motivated ultra-violet (UV) complete BSM theory, simplified models or effective theories can be useful tools to study potential deviations from SM predictions in a generic way.
An example of this is the effective $Z'$-model with new gauge boson couplings to the SM fermions, as discussed e.g. in~\cite{ParticleDataGroup:2022pth,Leike:1998wr,Langacker:2008yv}.
The idea is that an experimental determination of such couplings could eventually give guidance toward a specific UV-complete model which in turn might help explain other observational phenomena.

Recent measurements at the LHC have excluded the existence of a $Z'$ boson with a mass up to about 5\,TeV~\cite{ATLAS:2016gzy, ATLAS:2019fgd, ATLAS:2019erb, CMS:2019gwf, ATLAS:2020fry, CMS:2021ctt, CMS:2021klu,Alvarez:2020yim, Lozano:2021zbu}, which will be raised to close to 8 TeV~\cite{ATLAS:2018tvr}. Several ways to search for neutral gauge bosons have also been considered for future $e^+e^-$ colliders which, mainly due to a cleaner collision environment,
could probe masses up to about 20\,TeV, depending on the collider energy~\cite{Cvetic:1995zs, Leike:1996pj, Leike:1998wr, Godfrey:2005pm, Osland:2009dp, Andreev:2012cj, Han:2013mra, Godfrey:2013eta, Kapukchyan:2013hfa, Pankov:2017dkv, Gulov:2017uqa}. Given recent progress in the development of the muon collider accelerator design, the possibility of searching for heavy particles at such a multi-TeV machine is an appealing idea and has already been considered in \cite{Lu:2023jlr} where several resonance production channels have been studied.

A multi-TeV muon collider has been proposed as the next energy-frontier collider in high-energy physics, combining the clean environment of $e^+e^-$ machines with the energy reach of hadron colliders. Recent technological progress, especially in muon cooling~\cite{MuonCollider:2022glg}, has led to a high priority in the US Snowmass Community Summer Study~\cite{MuonCollider:2022ded, MuonCollider:2022nsa,Aime:2022flm,MuonCollider:2022xlm} and the Particle Physics Project Prioritization Panel (P5) report~\cite{P5report}. In Europe, the International Muon Collider Collaboration (IMCC) is leading the European effort, suggested in the last European Particle Physics Strategy Update (EPPSU 2020)~\cite{Accettura:2023ked,EuropeanStrategyGroup:2020pow}.

A comprehensive, systematic study of the indirect discovery reach of future muon colliders for heavy neutral gauge bosons originating from different unification scenarios has been missing in the literature. We address the issue in this paper by proposing a framework based on leptonic observables suitable not only for searches for a $Z'$ boson but also for specifying its nature. The work is structured as follows: in Section~\ref{sec:theory_intro}, we review models introducing additional neutral gauge bosons and in Section~\ref{sec:exp_intro}, we explain why future muon colliders may be used to search for them. In Section~\ref{sec:analysis}, we present our analysis procedure whose results are discussed in Section~\ref{sec:results}. The most important findings of the work are summarized in Section~\ref{sec:conclusions}.

\section{Theoretical background}
\label{sec:theory_intro}

In this section, we present the models considered in our study which contain additional neutral gauge bosons beyond the SM and review the relevant aspects of their phenomenology. Such a collection can of course never be complete, but we give a well-defined selection of different kinds of weakly and strongly-coupled models to cover theory space as comprehensively as possible.

\subsection{Gauge bosons beyond the Standard Model}
The Standard Model can be formulated as the most general renormalizable quantum field theory invariant under the gauge group $\mathcal{G}_{SM} = SU(3)_c \times SU(2)_L \times U(1)_Y$ with the matter content discovered as of today.
The symmetry under this group gives rise to the existence of the well-known gauge bosons: $g, \, W^\pm, \, Z, \, \gamma$.
Leaving gravity aside, it remains an open question whether $\mathcal{G}_{SM}$ represents the complete symmetry group responsible for what we can observe in Nature.
It is possible that it merely represents the broken version of an enlarged symmetry group
or \textit{Grand Unified Theory} (GUT)~\cite{Georgi:1974sy,Pati:1974yy,Gell-Mann:1979vob,Slansky:1981yr} which in turn could lead to the existence of additional gauge bosons among other new particles.
In general, any extension of the SM involving the introduction of additional gauge symmetries might extend the gauge boson sector~\cite{Leike:1998wr}.
The rich phenomenology of such models, reviewed for example in \cite{Langacker:2008yv}, could be connected to flavor non-universality~\cite{Langacker:2000ju, Baek:2006bv}, neutrino masses~\cite{Ma:1995xk, Keith:1996fv, Barger:2003zh, Kang:2004ix, King:2005jy, Mohapatra:1974gc} or the dark-matter problem~\cite{deCarlos:1997yv, Nakamura:2006ht, Barger:2007nv, Lee:2007mt, Belanger:2007dx}.

Since no direct hint towards a single specific model with enlarged gauge symmetry has been experimentally found to date, we study a variety of models in a general way, namely in terms of the phenomenology of additional heavy neutral gauge bosons they would ensue.
Therefore, we use an effective $Z'$-model 
where we introduce a $Z'$ with couplings to the SM fermions that are the only trace of New Physics.
In this model, the interactions of the neutral gauge bosons with the SM fermions are given by the generic neutral-current Lagrangian: \cite{Hewett:1988xc,Leike:1998wr,Langacker:2008yv,Gulov:2017uqa}:
\begin{eqnarray}
-\mathcal{L}_{NC}&=&e A_{\mu} J_{A}^\mu + g_Z Z_{\mu} J_{Z}^\mu +g_{Z'} Z_{\mu}^\prime J_{Z'}^\mu,
\nonumber\\
J_{A}^\mu &=& \sum_f \bar f \gamma^\mu  q_f f,
\nonumber\\
J_{Z}^\mu &=& \sum_f \bar f \gamma^\mu  (v_f^{SM}-\gamma_5a_f^{SM}) f,
\nonumber\\
J_{Z'}^\mu &=& \sum_f \bar f \gamma^\mu  (v_f-\gamma_5a_f) f,
\label{lag}
\end{eqnarray}
where $f$ denotes the SM fermions, $A$ is the photon, $Z$ is the SM neutral gauge boson, $Z'$ is the new heavy neutral boson, $e$ is the positron charge, $g_Z$ and $g_{Z'}$ are the gauge couplings of the corresponding bosons, $q_f$ is the fermion electric charge in units of $e$, $v_f^{SM}$ and $a_f^{SM}$ are the vector and axial-vector coupling of the fermion to the SM neutral boson and $v_f$ and $a_f$ are the couplings to the new neutral boson. Note that the splitting of the couplings into a prefactor $g_{Z'}$ and vector-/axial-vector components $v_f,a_f$, while customary for the SM, is rather arbitrary for a generic $Z'$ model. Our convention will become clear below.

\subsection{The set of $Z'$-models considered}

In our analysis, we consider a vast set of BSM models, including the Sequential Standard Model (denoted as SSM)~\cite{Barger:1980dx,Robinett:1981yz,Altarelli:1989ff}, the E6 Model (E6)~\cite{achiman1978quark,London:1986dk}, the Left-Right Symmetric Model (LR)~\cite{Pati:1974yy, Mohapatra:1974gc, Mohapatra:1979ia, Mohapatra:1980yp}, the Alternative Left-Right Model (ALR)~\cite{PhysRevD.36.274,Ashry_2015}, the Littlest Higgs Model (LH)~\cite{Arkani_Hamed_2002,Han_2003,Kilian:2003xt}, the Universal Simplest Little Higgs Model (USLH)~\cite{Schmaltz_2004,Kilian:2004pp,Kilian:2006eh}, and the $U(1)_X$ Model~\cite{Oda:2015gna,iso2009classically}. 
While the SSM cannot be embedded into a renormalizable GUT-like model, it has become a standard candle for $Z'$ searches with over-optimistically large couplings; for this reason, we keep it in our selection of models. Many early GUT models that were inspired by an embedding into a low-energy superstring action exhibit $E_6$ as a GUT gauge group. Several breaking scenarios via intermediate groups like $SU(5)$~\cite{Georgi:1974sy}, $SO(10)$~\cite{Fritzsch:1974nn}, Pati-Salam $SU(4)\times SU(2) \times SU(2)$~\cite{Pati:1974yy}, or trinification $SU(3)^3$~\cite{Glashow:1984gc}  have been studied, and the rank reduction from six to four for the SM gives rise to two potential $Z'$ candidates. The simplest class of models is left-right (LR)-symmetric models, which appear as low-scale effective theories of Pati-Salam or trinification likewise. Multi-step breaking with several GUT scales has been considered~\cite{Kilian:2006hh,Braam:2010sy}, and so have the effects of mixing between different $Z'$ bosons, both on model discrimination and reconstruction of the UV-scale model~\cite{Braam:2011xh,Rizzo:2012rf}. Besides weakly-coupled, GUT-inspired models, there is the large class of models of compositeness or partial compositeness where some of the additional symmetries in the strongly-coupled sector have been gauged to avoid too light or massless Nambu-Goldstone models. One class of models is Little Higgs models, while another class of strongly-coupled models via their dual description to warped extra dimensions~\cite{Randall:1999ee} directly interpolates into models with additional space-time dimensions like e.g. universal extra dimensions~\cite{Appelquist:2000nn,Han:1998sg}.

For simplicity, we assume that all the considered models are flavor-universal, \textit{i.e.}~the couplings do not differ between the three fermion generations. The scenarios considered are listed in Table~\ref{table:couplings} where we present the respective axial and vector couplings of the $Z'$  to the fermions together with the absolute normalization of the couplings, $g_{Z'}$.
For the $E_6$ models, the specific values of the 
mixing angle $\beta =0$, $\pi/2$, $\arctan(-
\sqrt{5/3})$ correspond to the so-called $\chi$, $\psi$ and $\eta$ models. We use $\beta =0$ in our
analysis as a representative case. For the LR
model, the mixing angle between the two $SU(2)$
groups has to be in the range
$\sqrt{2/3}\le\alpha\le\sqrt{c_W^2/s_W^2-1}$, and
we use the upper bound for $\alpha$. For the LH
model, the mixing angles between the two $SU(2)$
groups obey $1/10 \le c/s \le 2$,
and we use $c/s\equiv 1$.

\begin{table}
\centering
	\begin{tabular}[t]{lccc} 
		\toprule
		Model & $g_{Z'}$ & $2 v_l$ & $2 a_l$ \\ \midrule
		SSM & $\frac{e}{s_W c_W}$ & $2s_W^2-\frac{1}{2}$ & -$\frac{1}{2}$ \\ 
		$E_6$ & $\frac{e}{c_W}$ & $\frac{2\cos \beta}{\sqrt 6}$ & $\frac{\cos \beta}{\sqrt 6}+\frac{\sqrt {10}\sin \beta }{6}$ \\ 
		LR & $\frac{e}{c_W}$  & ${\frac{1}{\alpha}}- {\frac{\alpha}{2}}$ & ${\frac{\alpha}{2}}$ \\ 
		ALR & $\frac{e}{s_W c_W\sqrt{1-2 s_W^2}}$ & ${\frac{5}{2}s_W^2-1}$ & ${-\frac{1}{2}s_W^2}$ \\ 
		LH & $\frac{e}{s_W}$ & $-\frac{c}{4s}$ & $-\frac{c}{4s}$ \\ 
		USLH & $\frac{e}{c_W\sqrt{3-4 s_W^2}}$& $\frac{1}{2}-2s_W^2$ & $\frac{1}{2}$ \\ 
		$U(1)_X$ & $\frac{e}{4c_W}$ & -8 & 2\\
		\bottomrule 
	\end{tabular}
\caption{The $Z'$ couplings to leptons in the models considered in this paper. The sine of the Weinberg angle is denoted by $s_W$. For the description of the models and the explanation of the model-specific parameters see the main text. 
The table and notation are adapted from \cite{Gulov:2017uqa}.}
\label{table:couplings} 
\end{table} 

\subsection{Observables}
\label{sec:observables}
The models under consideration specify different points in the parameter space of axial and vector couplings to fermions.
Therefore, we study observables which encode angular and chiral properties in order to distinguish between the different scenarios.
The analysis is particularly simple in leptonic scattering channels because all of the models considered are flavor universal so that only two coupling factors enter the scattering amplitudes (as opposed to the production of hadrons where leptonic and hadronic axial and vector couplings would play a role, and couplings of up- and down-type quarks would be overlaid).
This makes it possible to study the models in the two-dimensional $a_l$-$v_l$ plane with the only other unknown being the mass of the $Z'$ (in principle, another additional parameter would be the $Z'$ width, but we will only consider $Z'$ states heavier than the collider energy such that the width does not have a significant impact on any observable).
The experimental observables we consider are:

\begin{enumerate}
\item the total cross-section for the process $\mu^{+}\mu^{-} \rightarrow f \bar{f}$, denoted as $\sigma^{f}$, for $f \in \{ e, \tau\}$ (we do not consider $\mu$ final states due to the contamination from $t$-channel exchange); 

\item  the forward-backward asymmetry, defined as:
\begin{equation}
    A^{f}_{FB} = \frac{\sigma^{f}_{F} - \sigma^{f}_{B}}{\sigma^{f}} ,
\end{equation} where, for $f \in \{ e, \tau \}$:
\begin{itemize}
    \item $\sigma^{f}_{F}$ -- the partial cross section for the fermion $f$ going in the forward direction,
    \begin{equation}
    \sigma^{f}_{F} = \int_0^1 d\cos\theta \frac{d\sigma}{d\cos\theta} (\mu^-\mu^+ \to f\bar{f})
    \end{equation}
    \item $\sigma^{f}_{B}$ -- the partial cross section for the fermion $f$ going in the backward direction;
    \begin{equation}
    \sigma^{f}_{B} = \int_{-1}^0 d\cos\theta \frac{d\sigma}{d\cos\theta} (\mu^-\mu^+ \to f\bar{f})
    \end{equation}
\end{itemize}

\item  the left-right asymmetry, defined as:
\begin{equation}
\label{eq:left-right}
    A_{LR}^{f} = \frac{\sigma^{f}_{LR} - \sigma^{f}_{RL}}{\sigma^{f}},
\end{equation}
where, for $f \in \{ e, \tau \}$:
\begin{itemize}
    \item $\sigma^{f}_{LR} = \sigma(\mu^-_L\mu^+_R \to f\bar{f})$ -- the partial cross section for the fully left-polarized muon beam and right-polarized antimuon beam,
    \item $\sigma^{f}_{RL} = \sigma(\mu^-_R\mu^+_L \to f\bar{f})$ -- the partial cross section for the fully right-polarized muon beam and left-polarized antimuon beam;
\end{itemize}
We will comment on the possibility of polarization at the muon collider below.

\item the polarization asymmetry, defined as:
\begin{equation}
    A_{pol}^{f} =\frac{\sigma^{f}_{lh} - \sigma^{f}_{rh}}{\sigma^{f}},
\end{equation}
where, for $f = \tau$:
\begin{itemize}
    \item $\sigma^{f}_{lh} = \sigma(\mu^-\mu^+ \to \tau^-_L\tau^+_R)$ -- the partial cross section for the left-handed fermions in the final state,
    \item $\sigma^{f}_{rh} = \sigma(\mu^-\mu^+ \to \tau^-_R\tau^+_L)$ -- the partial cross section for the right-handed fermions in the final state.
\end{itemize}
\end{enumerate}
Note that for massless fermions, $A_{pol}^{f} = A_{LR}^{f}$ holds \cite{Leike:1998wr} but we keep them as two independent quantities due to their vastly different experimental measurement.
This will manifest itself in our statistical analysis via different systematic uncertainties.

In accelerator physics, it is not possible to fully polarize lepton beams and \eqref{eq:left-right} should be corrected by a factor of the effective polarization
\begin{equation}
\label{eq:Peff}
    P_\text{eff} = \frac{P^+ + P^-}{P^{+}P^{-} + 1},
\end{equation}
where $P^-$ ($P^+$) is the polarization fraction for the (anti-) muon beam and we assume opposite polarizations for the two beams.
Analogously, fully perfect flavor tagging and $\tau$ polarization measurements are not possible and their efficiencies contribute to the systematic uncertainty of the study. Since the Muon Collider project is still in a preliminary phase, we tackle the issue by postulating an overall systematic uncertainty to 1\% which roughly matches the order of magnitude of the statistical uncertainty. The polarization asymmetry is an exception though, because its uncertainty within the LEP measurements is much larger than other systematic uncertainty~\cite{ALEPH:2001uca, DELPHI:1999yne}. 
Therefore,  we conservatively assume a systematic uncertainty of $5\%$ here.
The impact of assuming different values is briefly discussed in~\ref{App:RP2}.

In the current scope of the analysis, we ignore additional information from studying hadronic observables. Their usage provides valuable input for the model discrimination; however, in view of the leptonic production channel at the muon collider, disentangling different products of couplings entering the expressions of our observables, including tagging and separating light up- and down-quark flavors, requires additional assumptions in our statistical analysis. This is beyond the scope of the present study. There is recent progress in tagging light-quark flavors using final-state QED radiation (see e.g. \cite{talk_Epiphany}), which would enable to partially disentangle these light-quark flavor couplings.
We will extend the study towards hadronic observables in the future.

\section{High-energy muon collider setup}
\label{sec:exp_intro}

For a sustainable future in high-energy collider physics, long-term planning is unavoidable. 
The high-luminosity upgrade of the Large Hadron Collider (HL-LHC) is now approved~\cite{ZurbanoFernandez:2020cco} which, after a decade of running, could run contemporarily with the start-up phase of or will be followed by an $e^+e^-$ Higgs factory~\cite{EuropeanStrategyGroup:2020pow, P5report,ILC:2013jhg,Behnke:2013lya,FCC:2018evy,CEPCStudyGroup:2023quu,Aicheler:2012bya,Linssen:2012hp}.
On a 20-year time scale, ambitious projects for parton collisions at energies of $\sim 10\ \text{TeV}$ could be realized either with a 100-TeV hadron machine~\cite{FCC:2018vvp,FCC:2018byv, Tang:2015qga} or a multi-TeV muon collider~\cite{Accettura:2023ked}.
Since the latter has recently regained attention due to the successful demonstration of the muon-cooling principles~\cite{MICE:2019jkl}, we decided to consider this proposal in our study, as already alluded to in the introduction.

Muon colliders would offer a versatile environment for both precision studies of SM and BSM phenomenology and high-energy searches, including the possible occurrence of new physics in electroweak interactions.
Muons can be efficiently accelerated in circular machines, as they are more than 200 times more massive than electrons which significantly reduces bremsstrahlung.
On the other hand, contrary to protons, muons are elementary, point-like particles offering a much cleaner collision environment\footnote{Note, however, the electroweak partonic content inside muons when applying a collinear factorization picture to high-energy collisions~\cite{Chen:2016wkt,Han:2020uid,Garosi:2023bvq}}.
Nevertheless, their finite lifetime poses a challenge for the design of both the accelerator complex and the detector.
For brevity, we will simplify the discussion of the experimental effects in the following, assuming global systematic uncertainties only and leaving the meticulous study of the experimental conditions
to the time when the final detector designs are available.

The muon collider community currently aims at achieving a collision energy of 10\,TeV with a future machine~\cite{Accettura:2023ked}.
An initial stage of 3\,TeV is foreseen on the path towards the targeted energy and possible extensions of the project are not excluded if technology permits.
As for the current design of the accelerator complex, the integrated luminosity, $\mathcal{L}_\text{int}$, scales with the square of the collision energy, $E_\text{CM}$.
The results presented below scale mostly trivially with the integrated luminosity, and different running scenarios can be easily deduced.
In our analysis, we assume that a 10-TeV muon collider would deliver 10 ab$^{-1}$ of data which can be extrapolated to other energies by taking:
\begin{equation}
    \mathcal{L}_\text{int} (E_\text{CM}) = 10\,\text{ab}^{-1} \left(\frac{E_\text{CM}}{10\,\text{TeV}}\right)^2.
    \label{eq:Lint}
\end{equation}

The default setup of the muon collider does not assume polarization of the muon beams. However, due to the production of muons from pion decays there is a certain level of polarization inherent in the beams, and circular lepton colliders automatically build up transverse polarization. This could be converted into longitudinal polarization using spin rotators which has been discussed in the technical accelerator reports. Therefore, we will also consider the possibility of 30\% polarization of both beams.

\section{Analysis procedure}
\label{sec:analysis}

In this section, we present our analysis procedure. We demonstrate a statistical framework that we use to set limits on the $Z'$ masses which could be probed at a future muon collider and show how one can distinguish between different models of New Physics.
We carry out the analysis in the Born approximation which was shown to give reliable results in off-peak regions of the relevant observables, as long as appropriate kinematic cuts on photon radiation are applied \cite{Leike:1998wr}.

\subsection{Mass reach}
Our statistical analysis is based on the $\chi^2$-test statistic:
 \begin{equation}
 \label{eq:chisq}
        \chi^2 (a,v,M_{Z'}) = \sum_{i=1}^{n_{ob}}\left[\frac{O_i^{\text{SM}} - O_i (a,v,M_{Z'})}{\Delta O_i^\text{SM}}\right]^2 + n_{ob},
\end{equation}
where $n_{ob}$ is the number of observables used, $O_i^{\text{SM}}$ is the value of the $i$-th observable predicted by the SM, $O_i(a,v,M_{Z'})$ is the value of the observable in a given model (defined uniquely by a pair of axial and vector couplings, $(a,v)$, and the $Z'$ mass, $M_{Z'}$)\footnote{For brevity, we will drop the parameters whenever there is
no danger of confusion.} and $\Delta O_i^\text{SM}$ is the measurement uncertainty, $\Delta O_{i}^\text{SM} =  \sqrt{ \Delta O_{i,\text{stat}}^2 + \Delta O_{i,\text{sys}}^2}$ for $\Delta O_{i,\text{stat}}$ and $\Delta O_{i,\text{sys}}$ being the statistical and systematic uncertainties, respectively.
The term ``$+ n_{ob}$'' in~\eqref{eq:chisq} comes about via fluctuations of experimental values around the theory expectations of the $O_i^\text{SM}$.
We show in \ref{App:chi2} that this simple approach perfectly coincides with the procedure of performing hundreds of pseudo-experiments to mimic real measurements by extracting observables from normal distributions.
That this is indeed a meaningful framework can also be seen by assuming that a given model shows no significant discrepancy from the SM (\textit{i.e.}~the difference between all the observables is zero by construction) which also means the first term gives no contribution to the total value and the true expected value of the $\chi^2$ distribution can only be restored by the second term.

In the following, we will assume that the statistical uncertainties are given by~\cite{Leike:1996pj}:
\begin{subequations}
  \begin{align}
    \Delta{\sigma^f} &= \frac{\sigma^f}{\sqrt{N_f}}, 
    \\
    \Delta A_{FB}^f &= \sqrt{\frac{1-\left(A^f_{FB}\right)^2}{N_f}}, 
    \\
    \Delta A^f_{LR} &= \sqrt{\frac{1-\left(P_\text{eff}A^f_{LR}\right)^2}{N_fP_\text{eff}^2}}, 
    \\
    \Delta A_{pol}^f &= \sqrt{\frac{1-\left( A^f_{pol} \right)^2}{N_f}}, 
\end{align}
\label{eq:stat_err}
\end{subequations}
where $P_\text{eff}$ is the polarization fraction of the initial state lepton, as defined in \eqref{eq:Peff}, and $N_f = \mathcal{L}_\text{int} \cdot \sigma^{f}$ is the number of expected events.

Given the design status of a future muon collider, a complete assessment of the systematic uncertainties is currently not possible.
Therefore, for sake of simplicity of the analysis, we will assume that the measurement is statistically limited and the corresponding systematic uncertainties do not exceed significantly the statistical error.
Thus, our analysis may be perceived as a hint of what the desired detector performance should be. In \ref{App:RP2}, we show the impact of varying the systematic errors.

We assume that a model gives predictions distinct from the SM if the value of the $\chi^2$ test exceeds the critical value at the confidence level of 95\% for the given number of degrees of freedom, $n_\text{d.o.f.}$, $\chi^2(a,v,M_{Z'}) > \chi^2_\text{crit}(n_\text{d.o.f.})$. In our analysis, we combine seven observables, namely:
\begin{itemize}
\itemindent=9pt
    \item[1-2.] the total cross section, $\sigma^f$,  for $f \in \{ e, \tau\}$,
    \item[3-4.] the forward-backward asymmetry, $A^f_{FB}$, for $f \in \{ e, \tau \}$,
    \item[5-6.] the left-right asymmetry, $A^f_{LR}$, for $f \in \{ e, \tau \}$,
    \item[7.] the polarization asymmetry, $A^f_{pol}$, for $f = \tau $,
\end{itemize}
as defined in Sec.~\ref{sec:observables}.
For the mass reach, we have $\chi^2_\text{crit}(n_\text{d.o.f.}=7) = 12.02$ which corresponds to the $90\%$-quantile because it is a one-sided test, while for the resolution power discussed in the next section, we have $\chi^2_\text{crit}(n_\text{d.o.f.}=7) = 14.07$.

\subsection{Resolution power}

The resolution power measures the compatibility of a fictitious measurement of a parameter pair $(a,v)$ for a fixed $M_{Z'}$ with a given reference $Z'$-model with couplings~$(a_\text{model},v_\text{model})$. Should the measurement of the couplings fall outside the region where $\chi^2_\text{model} < \chi^2_\text{crit}(n_\text{d.o.f.})$, we can distinguish it from the theoretical prediction and thus, exclude the given model at the $95\%$ confidence level. As in this case, our measurement is compared to a particular model, the test statistic given in~\eqref{eq:chisq} should be replaced by:
\begin{equation}
\label{eq:chisq2}
    \chi^2_\text{model} (a,v,M_{Z'}) = \sum_{i=1}^{n_{ob}}\left[\frac{O_i^{\text{model}} - O_i (a,v,M_{Z'})}{\Delta O_i^\text{model} }\right]^2 + n_{ob},
\end{equation}
where the only difference comes from the fact that the SM value of an observable is replaced by the observable predicted within the given model, $O_i^{\text{model}}$.

\section{Results}
\label{sec:results}

\begin{figure}
	\begin{center}
			\includegraphics[width=.6\columnwidth]{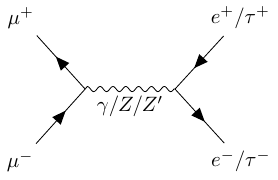}
	\end{center}
	\caption{The set of Feynman diagrams entering our $\chi^2$-analysis.}
	\label{fig:fd}
\end{figure}

The results presented in the following two subsections are obtained by determining values for each observable analytically using the full $(\mu^+ \mu^- \rightarrow \ell^+ \ell^-)$-scattering amplitudes for $\ell\in\{e,\tau\}$ in Born approximation shown in Figure~\ref{fig:fd}.
We also include the width of the $Z'$ using the Born approximations defined in~\cite{Leike:1998wr}.
The values for $\Gamma(Z'\to \text{any})$ range from $\sim 40\; \text{GeV}$ (USLH) to $\sim 300\; \text{GeV}$ (SSM) for $M_{Z'}=30\;\text{TeV}$.
The exact width of the $Z'$ does not play a major role in the analysis though, because the bounds we find are driven by the off-peak regions of the observables.

We use \eqref{eq:Lint} to set the integrated luminosity.
Then, the statistical errors of \eqref{eq:stat_err} are typically of $\mathcal{O}(1\%)$ or lower (for $M_{Z'} = 3 \sqrt{s}$).
This could serve as a target value for the systematic errors of the prospective collider measurements in order for the precision to be driven by statistical fluctuations rather than systematics. 

Cross-checks of the results obtained in this way were carried out in the Monte Carlo event generator framework Whizard~3.1~\cite{Kilian_2011, moretti2001omega}, using the included generic $Z'$-model implementation as well as its UFO interface~\cite{Christensen:2010wz,Degrande:2011ua,Darme:2023jdn}.
Note that the latter can in principle be used for more sophisticated event-level analyses, potentially also beyond the Born approximation~\cite{Bredt:2022dmm}, or involving detector simulation, but at the cost of higher computational effort. This is beyond the scope of this current study.

\subsection{Mass reach}
\begin{figure}
    \includegraphics[width=\linewidth]{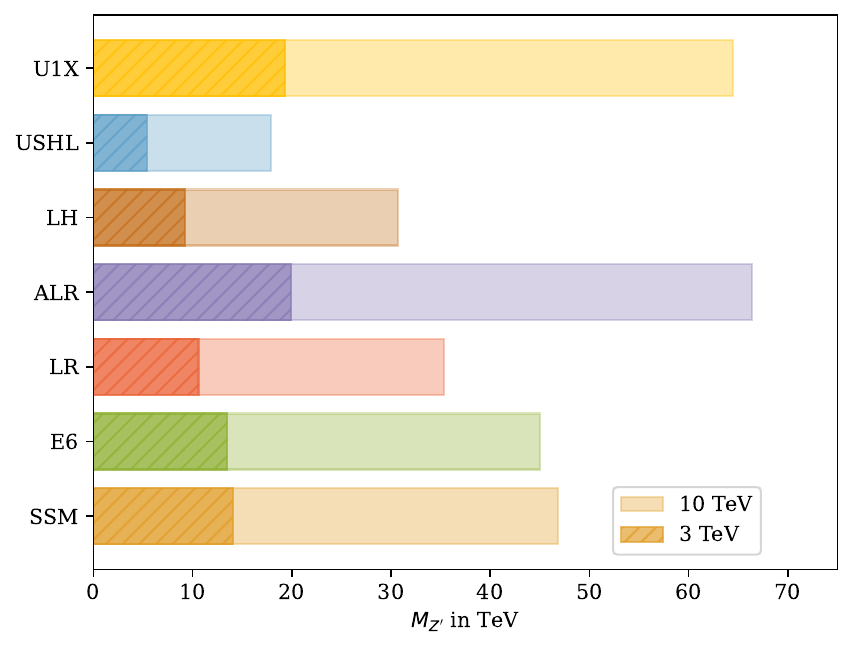}
	\caption{The reach in $M_{Z'}$ for a $3\,\text{TeV}$ and $10\,\text{TeV}$ muon collider with luminosities of $0.9\,\text{ab}^{-1}$ and $10\,\text{ab}^{-1}$, respectively for $P_\text{eff}=0$, $\Delta_{i,\text{sys}}=1\%$ for $\sigma_f$, $A_{FB}$, $A_{LR }$ and $\Delta_{A_{pol},\text{sys}}=5\%$.
	The bars correspond to the exclusion limit of the given $Z'$-model at 95\% confidence level.}
	\label{fig:reach}
\end{figure}
In Figure~\ref{fig:reach}, we show the mass reach we find using the analysis procedure explained in Sec.~\ref{sec:analysis}.
It shows that using leptonic observables alone, the exclusion limits at 95\% C.L.\ for a 10\;TeV muon collider extend up to $\sim\! 70\,\text{TeV}$, depending on the model.
We find relatively low exclusion limits of $\sim\! 17\;\text{TeV}$ for the Little/Littlest Higgs models due to the small magnitude of the leptonic axial and vector couplings\footnote{Note that in such a case also the current LHC bounds are weaker as the signal strengths in Drell-Yan searches is reduced.}.
The results shown are without beam polarization as the default setup of the muon collider, and because we find that even for polarization fractions close to $100\%$, the limits do not change significantly, \textit{i.e.}\ only by up to 3\%.
This is due to the limited statistical significance of $A_{LR}$, \textit{i.e.}~a higher statistical error compared to $A_{FB}$ (see \eqref{eq:stat_err}).
The influence of the polarization fraction becomes important for the model discrimination though, as explained in the following section.

The discussion of \cite{Leike:1996pj} for $e^+e^-$ colliders shows that the inclusion of hadronic observables can increase the reach by up to $\sim 50\%$, depending on the model.
An example is the SSM, where one has axial and vector couplings to up- and down-type quarks of relatively large magnitude.
The effect is much less drastic for models with smaller quark couplings though, such as the $U(1)_X$. For muon colliders, we expect a similar increase in mass reach when including hadronic observables.

\subsection{Resolution power}
\begin{figure*}
	\begin{subfigure}[b]{0.32\textwidth}
		\includegraphics[width=\linewidth]{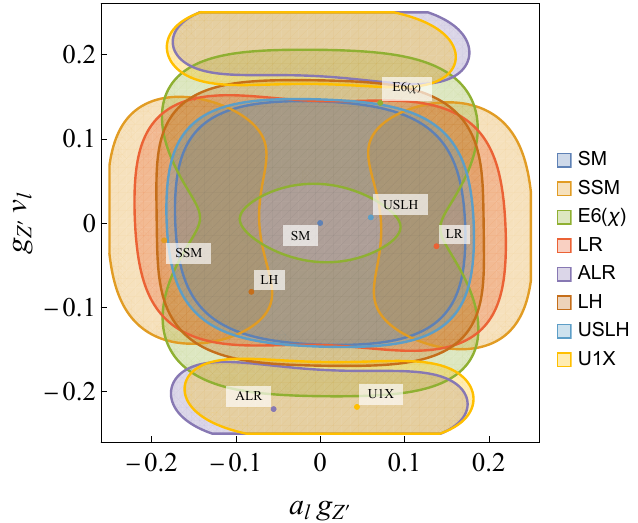}
		\caption{$M_{Z'} = 40\,\text{TeV}$}
		\label{fig:MZR40}
	\end{subfigure}%
	\begin{subfigure}[b]{0.32\textwidth}
		\includegraphics[width=\linewidth]{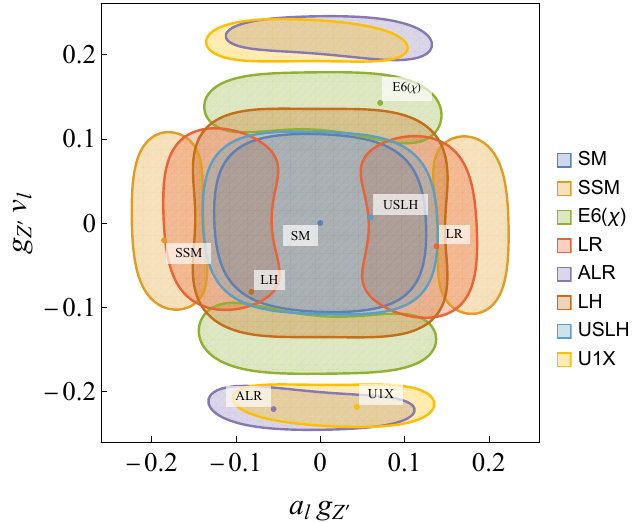}
		\caption{$M_{Z'} = 30\,\text{TeV}$}
		\label{fig:MZR30}
	\end{subfigure}%
 	\begin{subfigure}[b]{0.32\textwidth}
		\includegraphics[width=\linewidth]{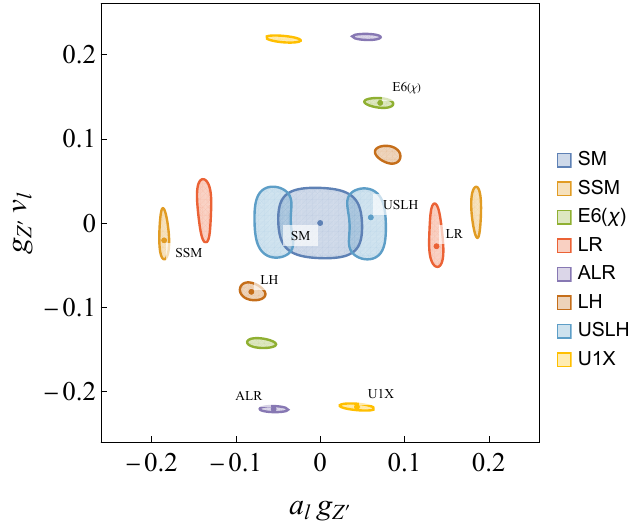}
		\caption{$M_{Z'} = 15\,\text{TeV}$}
		\label{fig:MZR15}
	\end{subfigure}%
	\caption{Resolution power for different masses of the $Z'$. Inputs: $\mathcal{L}_\text{int} = 10\,\text{ab}^{-1}$, $E_\text{CM} = 10 \,\text{TeV}$, $P_\text{eff}=0$.}
	\label{fig:MZR}
\end{figure*}
Here, we determine the resolution power of a 10\,TeV muon collider in the off-peak region of the observables considered, \textit{i.e.}\ for masses above the collider energy. A study in the peak region would require a different type of analysis and for even smaller masses, a study for precision $e^+e^-$ machine would be more sensible.
In Figure~\ref{fig:MZR}, we show the resolution power for $Z'$-axial and -vector couplings, given that a signal for a $Z'$ has been found at a mass $M_{Z'}$ of either 40\,TeV, 30\,TeV or 15\,TeV.
We find that one would be able to discriminate to a good extent between the models considered\footnote{Of course this does not comprise all conceivable Z' models in theory space, but rather gives an impression for the possibility of discriminating several benchmark scenarios via the measurement of low-energy effective couplings.} at around $M_{Z'} \sim 30\,\text{TeV} = 3\sqrt{s}$ or lower.
Already for masses slightly higher than that, the discrimination power starts to decrease rather rapidly.
On the other hand, it grows very fast for lower masses because the closer the $Z'$-mass is to the collider energy, the more one approaches the $Z'$-pole where deviations from the SM become substantial.
Then, models with axial or vector couplings of comparatively large magnitude can be discriminated even without the use of polarized beams or a very accurate measurement of the $\tau$ polarization. In \ref{App:RP2}, we discuss the influence of the latter measurements in more detail.

\section{Conclusions}
\label{sec:conclusions}
A $Z'$ boson occurs in a variety of BSM theories that are based on extensions of the SM gauge group. We showed that at a muon collider, using indirect search methods for leptonic observables alone, one would be able to probe $Z'$ masses of up to $\sim 70\,\text{TeV}$.
For masses up to $\sim 30\,\text{TeV}$, the same framework can be applied to discriminate its nature in terms of its axial and vector couplings to fermions, pointing towards a specific model of New Physics. The presented results give the most stringent limits up to date. Employing hadronic observables has been left for future explorations. They are expected to enhance the discovery reach towards higher $Z'$ masses but are not expected to improve the model discrimination by a lot, as they suffer from combinatorics between light up- and down-type quarks. Final states consisting of charm and bottom will add discrimination power but are more complicated to quantify due to the charm- and bottom-tagging efficiencies. $Z'$ decays into top quarks are highly interesting, especially for models where the top coupling is special (e.g. the Littlest Higgs or top-color models), but clearly these final states go beyond the simple two-fermion signatures considered here. Note that another complication is the non-factorization of leptonic production current and hadronic decay current.

A spectacular level of model discrimination is already possible using the default setup of the muon collider, using left-right and $\tau$ polarization asymmetries. If the systematic error on the $\tau$ polarization measurement dominates over its statistical uncertainty significantly, beam polarization of a degree of up to 30\% via spin rotators could enhance the discriminative power as it allows to use left-right asymmetries as well.

As statistical uncertainties can be expected to be of the order of 1\%, we show that it is crucial to have the systematic uncertainties under control in order to guarantee a good model discrimination. This means that systematic uncertainties from lepton charge determination, angular resolution, $\tau$ polarization measurements etc. should not exceed a percent by much.
This provides a ballpark what future muon collider detector development should aim for.

\section*{Acknowledgments}
The work of KM and JRR was partially supported by the National Science Centre
(Poland) under the OPUS research project no. 2021/43/B/ST2/01778.
ML, KM and JRR acknowledge the support of the Deutsche
Forschungsgemeinschaft (DFG, German Research Association) under
Germany's Excellence Strategy-EXC 2121 ``Quantum Universe''-390833306. KK and MM acknowledge Deutsches Elektronen-Synchrotron (DESY) for a hospitable environment during the DESY Ukraine Winter School. This work
has also been funded by the Deutsche Forschungsgemeinschaft (DFG,
German Research Foundation) -- 491245950. Furthermore, we acknowledge support from
the COMETA COST Action CA22130.

\appendix
\section{Parametric variations of resolution power}
\label{App:RP2}
In this Appendix, we present a list of plots for the discrimination power in the $a_l$-$v_l$ plane of $Z'$ models for variations of different input values.
\begin{figure*}
	\begin{subfigure}[b]{0.32\textwidth}
		\includegraphics[width=\linewidth]{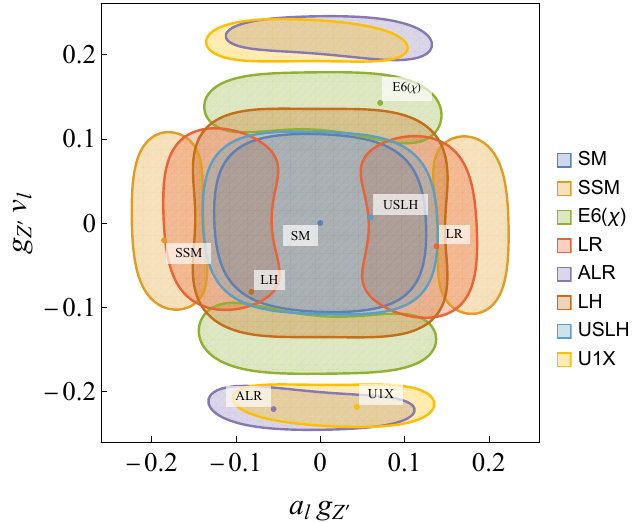}
		\caption{$P_\text{eff} = 0\%$, $\Delta_{\text{sys}}(A_{pol}) = 5\%$}
		\label{fig:RP0p5Err}
	\end{subfigure}%
	\begin{subfigure}[b]{0.32\textwidth}
		\includegraphics[width=\linewidth]{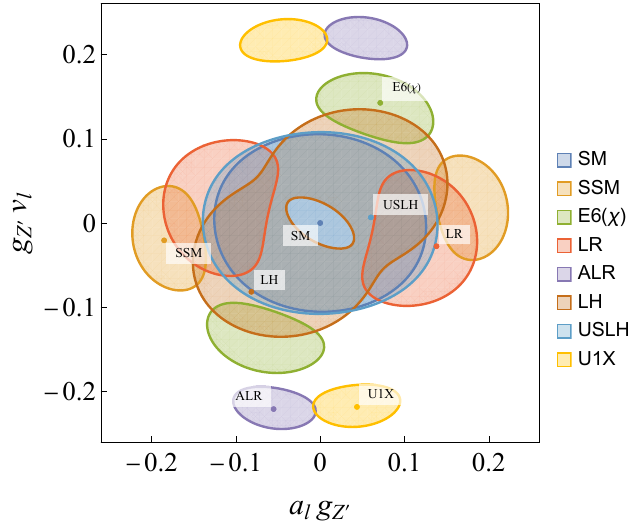}
		\caption{$P_\text{eff} = 0\%$, $\Delta_{\text{sys}}(A_{pol}) = 1\%$}
		\label{fig:RP0p1Err}
	\end{subfigure}%
	\begin{subfigure}[b]{0.32\textwidth}
		\includegraphics[width=\linewidth]{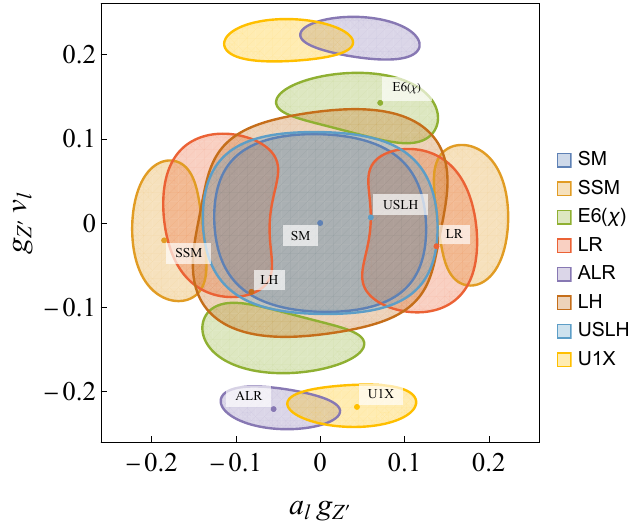}
		\caption{$P_\text{eff} = 30\%$, $\Delta_{\text{sys}}(A_{pol}) = 5\%$}
		\label{fig:RP03p5Err}
	\end{subfigure}%
	\caption{Resolution power for different combinations of effective polarizations and systematic errors for the $\tau$ polarization asymmetry. Inputs: $\mathcal{L}_\text{int} = 10\,\text{ab}^{-1}$, $E_\text{CM} = 10 \,\text{TeV}$, $M_{Z'}=30\,\text{TeV}$.}
	\label{fig:RP}
\end{figure*}
In Figure~\ref{fig:RP}, we show that adding a polarization measurement or polarized beams to the analysis enables the distinction of models with different relative signs between the axial and vector couplings, \text{e.g.}~between the ALR and $U(1)_X$ models. 
A simultaneous sign change of the axial and vector couplings would remain undetectable, because the left-right asymmetry $A_{LR}$ only adds sensitivity to the product $a_l \cdot v_l$ \cite{Leike:1998wr}.
Comparing Figure~\ref{fig:RP0p1Err} to Figure~\ref{fig:RP03p5Err}, we see that one can achieve similar resolutions by either bringing the systematic uncertainty on the $\tau$ polarization down to the 1\%-level or by using polarized muon beams with $P_\text{eff} = 30\%$. Moreover, we find that a further increase in the effective polarization does not yield a significant increase in the resolution power.
In \ref{App:RP}, we show how the individual observables contribute to the resolution power and present their analytic form in \ref{App:asym}.

\begin{figure*}
	\begin{subfigure}[b]{0.32\textwidth}
		\includegraphics[width=\linewidth]{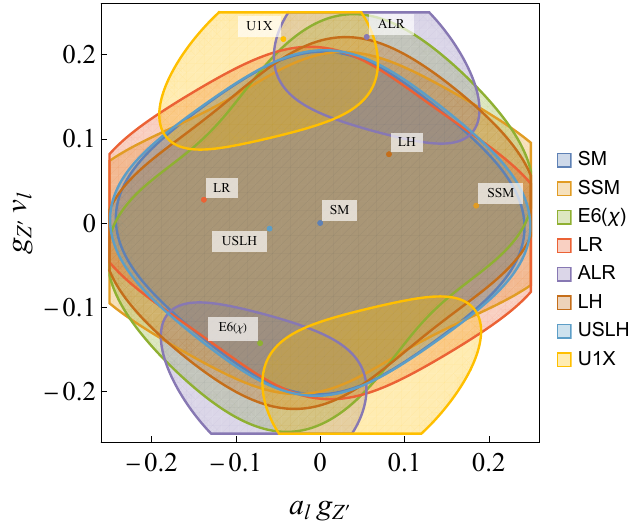}
		\caption{$\Delta_{\text{sys},i} = 5\%$}
		\label{fig:Err05}
   \end{subfigure}%
  	\begin{subfigure}[b]{0.32\textwidth}
		\includegraphics[width=\linewidth]{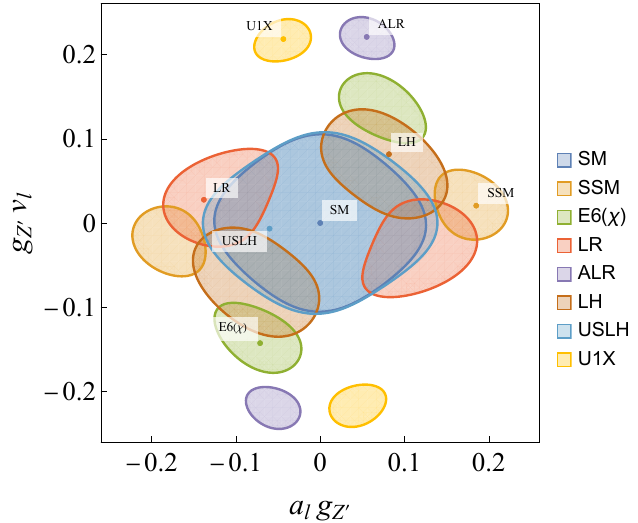}
		\caption{$\Delta_{\text{sys},i} = 1\%$}
		\label{fig:Err01}
	\end{subfigure}%
	\begin{subfigure}[b]{0.32\textwidth}
		\includegraphics[width=\linewidth]{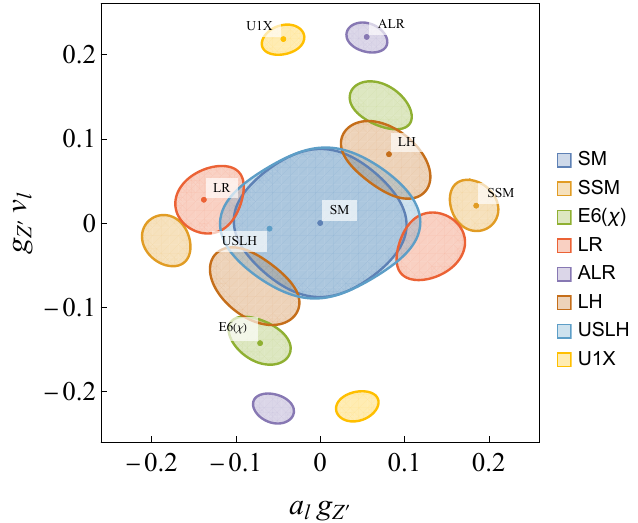}
		\caption{$\Delta_{\text{sys},i} = 0.1\%$}
		\label{fig:Err001}
	\end{subfigure}%
	\caption{Resolution power for different systematic errors. Inputs: $\mathcal{L}_\text{int} = 10\,\text{ab}^{-1}$, $E_\text{CM} = 10 \,\text{TeV}$, $M_{Z'}=30\,\text{TeV}$, $P_\text{eff}=80\%$.}
	\label{fig:Err}
\end{figure*}
In Figure~\ref{fig:Err} we show the variations for three different values of the systematic errors.
We find that at $\sim 1\%$, systematic errors become competitive both with the statistical ones by explicit comparisons of the errors magnitudes, as well as by noticing that even a strong reduction in the systematic error does not increase the discrimination power significantly, as then the measurement becomes limited by statistics.

We do not show the variation of other parameters explicitly here, because their influence can be inferred from the variations of the parameters discussed.
The variation of the collider energy yields similar changes as shown in Figure~\ref{fig:MZR}, namely that the closer the energy is to the mass of the $Z'$, the better the discrimination power will be.
A variation of the luminosity solely affects the magnitude of the statistical errors.
Therefore, a decrease in luminosity will lead to larger errors and similar behavior as shown in Figure~\ref{fig:Err}.

\section{Test statistics derivation}
\label{App:chi2}
In this section, we discuss the origin of the shift in our $\chi^2$-distributions by the number of observables, as it appears in \eqref{eq:chisq} and \eqref{eq:chisq2}.
Let us call the pseudo-measurement of an observable for a given reference model $\hat{O}_i$ and the corresponding theory prediction $O_i$ with the error $\Delta O_i$ and assume that they follow the normal distribution
\begin{equation}
    \label{eq:normdist}
    f(\hat{O}_i) = \exp\left[-\frac{(\hat{O}_i-O_i)^2}{\Delta O_i\,\sqrt{2\pi}}\right] \frac{1}{\Delta O_i\,\sqrt{2\pi}}.
\end{equation}
This means that the expectation value for $\hat{O}_i$ is
\begin{equation}
    \langle\hat{O}_i\rangle = O_i
\end{equation}
and the variance
\begin{equation}
    \langle(\hat{O}_i-O_i)^2\rangle = \Delta O_i^2.
\end{equation}
We can also expand the variance:
\begin{equation}
    \langle(\hat{O}_i-O_i)^2\rangle = \langle\hat{O}_i^2\rangle -2 \langle \hat{O}_i \rangle O_i + O_i^2 = \langle\hat{O}_i^2\rangle - O_i^2,
\end{equation}
and therefore rewrite:
\begin{equation}
\label{eq:exp-val-Oisq}
\langle\hat{O}_i^2\rangle = O_i^2 + \Delta O_i^2.
\end{equation}
If we now determine the expectation value for $\chi^2$ to test a model with observables $O_i^\text{test}$ and insert \eqref{eq:exp-val-Oisq}, we find
\begin{align}
    \langle\chi^2\rangle &= \sum_{i=1}^{n_{ob}}\left\langle\frac{(\hat{O}_i - O_i^\text{test})^2}{\Delta O_i^2}\right\rangle
    = \sum_{i=1}^{n_{ob}}\frac{(O_i - O_i^\text{test})^2}{\Delta O_i^2} + n_{ob},
\end{align}
where the shift by $n_{ob}$ appears due to the second term in \eqref{eq:exp-val-Oisq} and explains why it is included in \eqref{eq:chisq} and \eqref{eq:chisq2}.
This is equivalent to the outcome when averaging over a large number of pseudo-measurements when drawing values from \eqref{eq:normdist} randomly using a Monte Carlo algorithm.

\section{Resolution power of individual observables}
\label{App:RP}
\begin{figure*}
	\begin{subfigure}[b]{0.32\textwidth}
		\includegraphics[width=\linewidth]{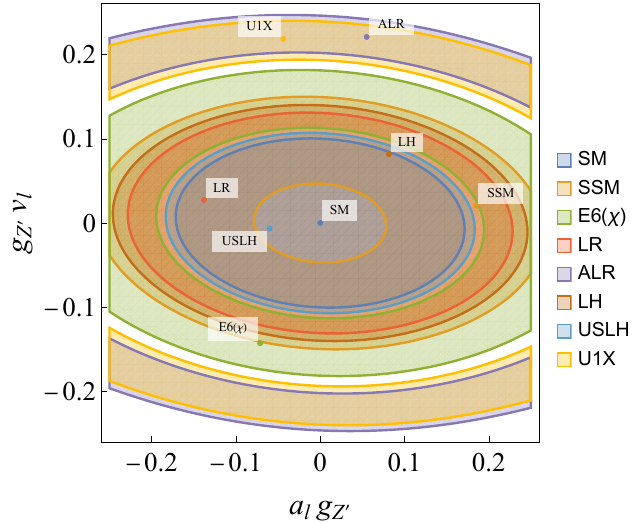}
		\caption{$\sigma_\text{tot}$}
		\label{fig:RPXsec}
	\end{subfigure}%
	\begin{subfigure}[b]{0.32\textwidth}
		\includegraphics[width=\linewidth]{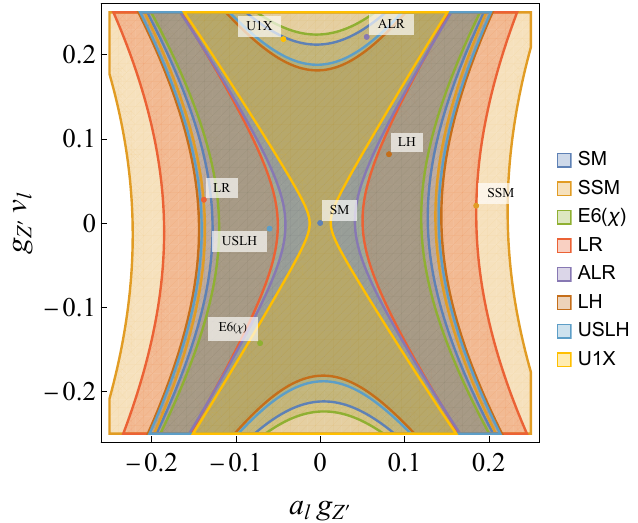}
		\caption{$A_{FB}$}
		\label{fig:RPAFB}
	\end{subfigure}%
	\begin{subfigure}[b]{0.32\textwidth}
		\includegraphics[width=\linewidth]{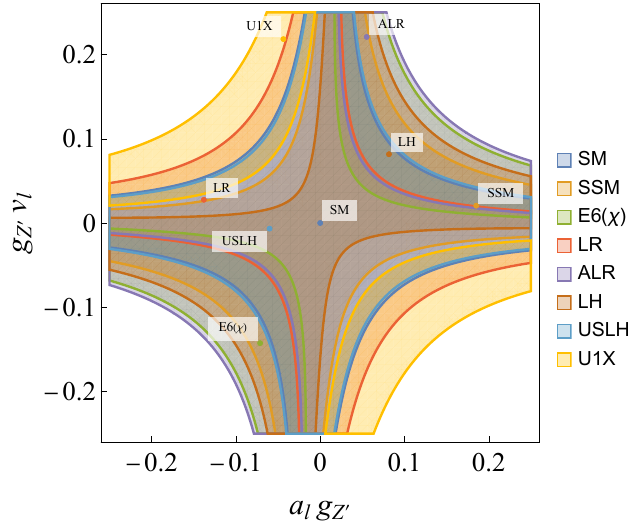}
		\caption{$A_{LR}$}
		\label{fig:RPALR}
	\end{subfigure}%
	\caption{Resolution power of individual observables. Inputs: $\mathcal{L}_\text{int} = 10\,\text{ab}^{-1}$, $E_\text{CM} = 10 \,\text{TeV}$, $M_{Z'}=30\,\text{TeV}$.}
	\label{fig:RPind}
\end{figure*}
In Figure~\ref{fig:RPind}, we show the resolution power of the individual observables $\sigma_\text{tot}$, $A_{FB}$ and $A_{LR}$.
The total cross section gives elliptic resolution regions while the two asymmetries give distinct hyperbolic regions.
This can be understood by the dependence of the observables on either the sum of the squares of axial and vector couplings ($\sigma_\text{tot}$), the difference of the squares ($A_{FB}$) or their product ($A_{LR}$) in the Born approximation \cite{Leike:1996pj}.
In comparison to the approximate bounds for the observables used in \cite{Leike:1996pj}, we see slight asymmetries in the bounding regions as a result of using the full Born amplitudes.
They are nevertheless in very good agreement with the approximate bounds.

\section{Asymmetries}
\label{App:asym}
\begin{figure*}
	\begin{subfigure}[b]{0.50\textwidth}
		\includegraphics[width=\linewidth]{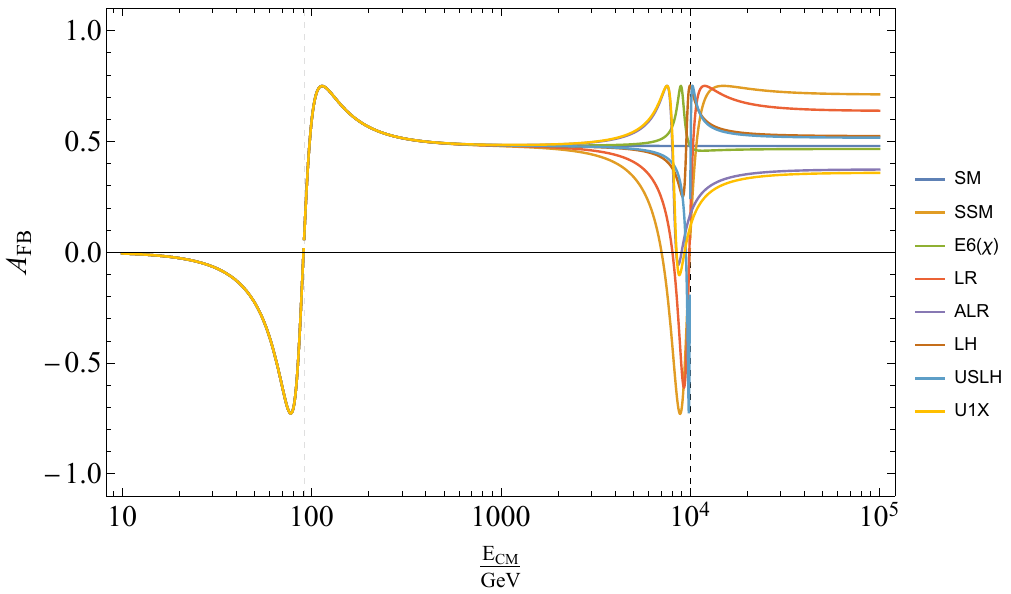}
		\label{fig:AFBana}
   \end{subfigure}%
  	\begin{subfigure}[b]{0.50\textwidth}
		\includegraphics[width=\linewidth]{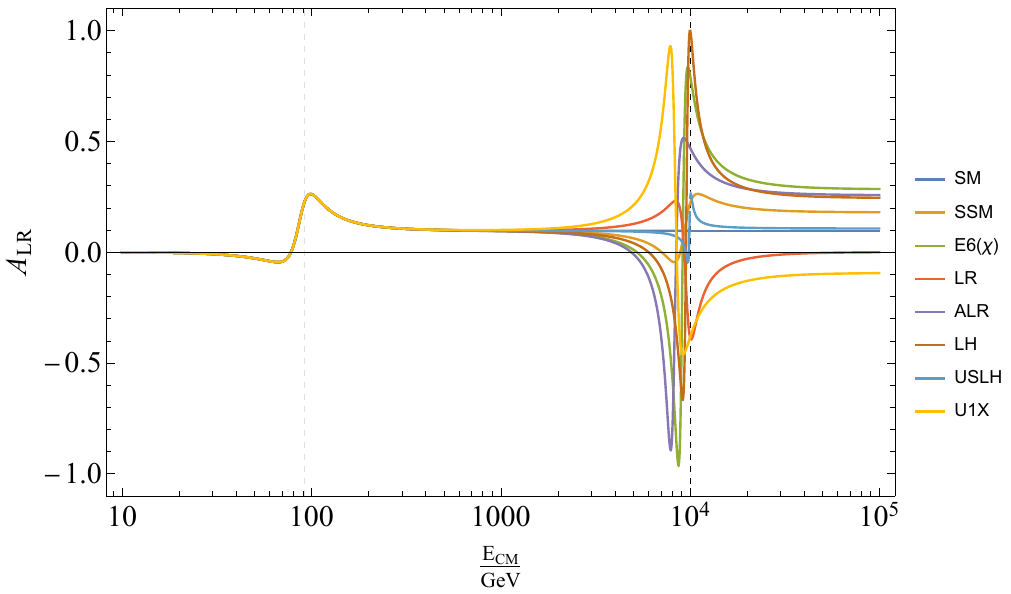}
		\label{fig:ALRana}
	\end{subfigure}%
	\caption{We show our results for the forward-backward and left-right asymmetry that enter our $\chi^2$-analysis as function of the c.o.m.\ energy. The dashed lines lie at the mass poles at $M_Z = 91.1882\,\text{GeV}$ and $M_{Z'} = 10\,\text{TeV}$.}
	\label{fig:asymmetries}
\end{figure*}
Here we present expressions for left-right and forward-backward asymmetries in terms of individual interference diagrams (see Fig.~\ref{fig:fd}) for a high-energy machine (\textit{i.e.}\ we neglect the masses of the fermions).
The forward-backward asymmetry reads
\begin{strip}
\rule[-1ex]{\columnwidth}{1pt}\rule[-1ex]{1pt}{1.5ex}
\begin{align}
    A_{FB}^{f}(s)=\frac{1}{8 \:\pi\: s\: \sigma^{f}(s) }\bigg\{& 2g_{Z}^4a^{SM}_{\mu}a^{SM}_{f}v^{SM}_{\mu}v^{SM}_{f}|\chi_Z|^2+2g_{Z'}^4a_{\mu}a_{f}v_{\mu}v_{f}|\chi_{Z'}|^2
    +q_\mu q_f e^2 \Big(g_{Z}^2 a^{SM}_{\mu} a^{SM}_{f}  \Re(\chi_\gamma \chi_Z^*)+g_{Z'}^2a_{\mu} a_{f}\Re(\chi_\gamma \chi_{Z'}^*)\Big) \nonumber \\
    &+g_{Z}^2 g_{Z'}^2(a^{SM}_{\mu} v_{\mu}+a_{\mu} v^{SM}_{\mu}) (a^{SM}_{f} v_{f}+a_{f} v^{SM}_{f}) \Re( \chi_Z \chi_{Z'}^*)\bigg\},
\end{align}
the left-right asymmetry is
\begin{align}
\begin{split}
    A_{LR}^{f}(s)=\frac{1}{12 \:\pi\: s\: \sigma^{f}(s) }\bigg\{
    &
    g_{Z}^4 \Big( a^{SM}_{\mu} v^{SM}_{\mu}\Big[(a^{SM}_{f})^2+(v^{SM}_{f} )^2\Big]+a^{SM}_{f} v^{SM}_{f} \Big[(a^{SM}_{\mu})^2+(v^{SM}_{\mu})^2\Big] \Big)|\chi_Z|^2 
    \\
    &
    +g_{Z'}^4\Big(a_{f} v_{f} \Big[a^{2}_{\mu}+v^{2}_{\mu}\Big] + a_{\mu} v_{\mu}\Big[a^{2}_{f}+ v^{2}_{f} \Big]\Big)|\chi_{Z'}|^2
    \\
    &
    + e^2 q_\mu q_f \Big[g_{Z}^2 (a^{SM}_{f} v^{SM}_{\mu} + a^{SM}_{\mu} v^{SM}_{f}) \Re(\chi_\gamma \chi_Z^*)
    +g_{Z'}^2(a_{f} v_{\mu} + a_{\mu} v_{f})  \Re(\chi_\gamma \chi_{Z'}^*)\Big]\\
    &+g_{Z}^2 g_{Z'}^2\Big[(a^{SM}_{\mu} a_{\mu}+v^{SM}_{\mu}v_{\mu})(a^{SM}_{f}v_{f}+v^{SM}_{f}a_{f})+(a^{SM}_{f} a_{f}+v^{SM}_{f}v_{f})(a^{SM}_{\mu}v_{\mu}+v^{SM}_{\mu}a_{\mu})\Big]\Re( \chi_Z \chi_{Z'}^*)\bigg\},\\
\end{split}
\end{align}
and for massless fermions, it holds that $A_{pol} = A_{LR}$.
The total cross section in this limit is given by
\begin{align}
\begin{split}
     \sigma^{f}(s)=\frac{1}{12 \pi s}  \bigg\{
     &
     g_{Z}^4\Big[(a^{SM}_{\mu})^2+(v^{SM}_{\mu})^2)\Big]\Big[(a^{SM}_{f})^2+(v^{SM}_{f})^2\Big]|\chi_Z|^2
     + g_{Z'}^4(a_{\mu}^2+v_{\mu}^2)(a_{f}^2+v_{f}^2)|\chi_{Z'}|^2+ e^4 q_\mu^2 q_f^2|\chi_\gamma|^2
     \\
     &
     +2 e^2 q_\mu q_f\Big[ g_{Z}^2 v^{SM}_{f} v^{SM}_{\mu}  \Re(\chi_\gamma \chi_Z^*)+ g_{Z'}^2 v_{f} v_{\mu} \Re(\chi_\gamma \chi_{Z'}^*)\Big]
     \\
     &
     +2 g_{Z}^2 g_{Z'}^2 \Big[a^{SM}_{f} a_{f}  +v^{SM}_{f} v_{f}\Big]\Big[a^{SM}_{\mu} a_{\mu}  +v^{SM}_{\mu} v_{\mu}\Big] \Re( \chi_Z \chi_{Z'}^*)\bigg\}.\\
\end{split}
\end{align}
\hfill\rule[1ex]{1pt}{1.5ex}\rule[2.3ex]{\columnwidth}{1pt}
\end{strip}
Here, $\chi_{Z^{(\prime)}}$ are the gauge bosons propagators times a factor of $s$ to make them dimensionless, \textit{i.e.} the ratio of the photon propagator to considered gauge boson,
\begin{align}
    \chi_{Z^{(\prime)}} = \frac{s}{s + i \Gamma_{Z^{(\prime)}} M_{Z^{(\prime)}} - M_{Z^{(\prime)}}^2 }\; ,
\end{align}
and $\chi_\gamma = 1$. This way of writing illustrates which interference of amplitude times conjugate amplitude is each term's origin when squaring the diagrams in Fig.~\ref{fig:fd}.
The distributions are shown in Fig.~\ref{fig:asymmetries}.
Note that the formulae above do not hold for final state muons because one would have additional $t$-channel contributions there.

\bibliographystyle{elsarticle-num}
\bibliography{bibliography.bib}

\begin{thebibliography}{10}
\expandafter\ifx\csname url\endcsname\relax
  \def\url#1{\texttt{#1}}\fi
\expandafter\ifx\csname urlprefix\endcsname\relax\def\urlprefix{URL }\fi
\expandafter\ifx\csname href\endcsname\relax
  \def\href#1#2{#2} \def\path#1{#1}\fi

\bibitem{ParticleDataGroup:2022pth}
R.~L. Workman, et~al., {Review of Particle Physics}, PTEP 2022 (2022) 083C01.
\newblock \href {https://doi.org/10.1093/ptep/ptac097}
  {\path{doi:10.1093/ptep/ptac097}}.

\bibitem{Leike:1998wr}
A.~Leike, {The Phenomenology of extra neutral gauge bosons}, Phys. Rept. 317
  (1999) 143--250.
\newblock \href {http://arxiv.org/abs/hep-ph/9805494}
  {\path{arXiv:hep-ph/9805494}}, \href
  {https://doi.org/10.1016/S0370-1573(98)00133-1}
  {\path{doi:10.1016/S0370-1573(98)00133-1}}.

\bibitem{Langacker:2008yv}
P.~Langacker, {The Physics of Heavy $Z^\prime$ Gauge Bosons}, Rev. Mod. Phys.
  81 (2009) 1199--1228.
\newblock \href {http://arxiv.org/abs/0801.1345} {\path{arXiv:0801.1345}},
  \href {https://doi.org/10.1103/RevModPhys.81.1199}
  {\path{doi:10.1103/RevModPhys.81.1199}}.

\bibitem{ATLAS:2016gzy}
M.~Aaboud, et~al., {Search for resonances in diphoton events at $\sqrt{s}$=13
  TeV with the ATLAS detector}, JHEP 09 (2016) 001.
\newblock \href {http://arxiv.org/abs/1606.03833} {\path{arXiv:1606.03833}},
  \href {https://doi.org/10.1007/JHEP09(2016)001}
  {\path{doi:10.1007/JHEP09(2016)001}}.

\bibitem{ATLAS:2019fgd}
G.~Aad, et~al., {Search for new resonances in mass distributions of jet pairs
  using 139 fb$^{-1}$ of $pp$ collisions at $\sqrt{s}=13$ TeV with the ATLAS
  detector}, JHEP 03 (2020) 145.
\newblock \href {http://arxiv.org/abs/1910.08447} {\path{arXiv:1910.08447}},
  \href {https://doi.org/10.1007/JHEP03(2020)145}
  {\path{doi:10.1007/JHEP03(2020)145}}.

\bibitem{ATLAS:2019erb}
G.~Aad, et~al., {Search for high-mass dilepton resonances using 139 fb$^{-1}$
  of $pp$ collision data collected at $\sqrt{s}=$13 TeV with the ATLAS
  detector}, Phys. Lett. B 796 (2019) 68--87.
\newblock \href {http://arxiv.org/abs/1903.06248} {\path{arXiv:1903.06248}},
  \href {https://doi.org/10.1016/j.physletb.2019.07.016}
  {\path{doi:10.1016/j.physletb.2019.07.016}}.

\bibitem{CMS:2019gwf}
A.~M. Sirunyan, et~al., {Search for high mass dijet resonances with a new
  background prediction method in proton-proton collisions at $\sqrt{s} =$ 13
  TeV}, JHEP 05 (2020) 033.
\newblock \href {http://arxiv.org/abs/1911.03947} {\path{arXiv:1911.03947}},
  \href {https://doi.org/10.1007/JHEP05(2020)033}
  {\path{doi:10.1007/JHEP05(2020)033}}.

\bibitem{ATLAS:2020fry}
G.~Aad, et~al., {Search for heavy diboson resonances in semileptonic final
  states in pp collisions at $\sqrt{s}=13$ TeV with the ATLAS detector}, Eur.
  Phys. J. C 80~(12) (2020) 1165.
\newblock \href {http://arxiv.org/abs/2004.14636} {\path{arXiv:2004.14636}},
  \href {https://doi.org/10.1140/epjc/s10052-020-08554-y}
  {\path{doi:10.1140/epjc/s10052-020-08554-y}}.

\bibitem{CMS:2021ctt}
A.~M. Sirunyan, et~al., {Search for resonant and nonresonant new phenomena in
  high-mass dilepton final states at $ \sqrt{s} $ = 13 TeV}, JHEP 07 (2021)
  208.
\newblock \href {http://arxiv.org/abs/2103.02708} {\path{arXiv:2103.02708}},
  \href {https://doi.org/10.1007/JHEP07(2021)208}
  {\path{doi:10.1007/JHEP07(2021)208}}.

\bibitem{CMS:2021klu}
A.~Tumasyan, et~al., {Search for heavy resonances decaying to WW, WZ, or WH
  boson pairs in the lepton plus merged jet final state in proton-proton
  collisions at $\sqrt{s}$ = 13 TeV}, Phys. Rev. D 105~(3) (2022) 032008.
\newblock \href {http://arxiv.org/abs/2109.06055} {\path{arXiv:2109.06055}},
  \href {https://doi.org/10.1103/PhysRevD.105.032008}
  {\path{doi:10.1103/PhysRevD.105.032008}}.

\bibitem{Alvarez:2020yim}
E.~Alvarez, M.~Est\'evez, R.~M. Sand\'a~Seoane, {Z'-explorer: A simple tool to
  probe Z' models against LHC data}, Comput. Phys. Commun. 269 (2021) 108144.
\newblock \href {http://arxiv.org/abs/2005.05194} {\path{arXiv:2005.05194}},
  \href {https://doi.org/10.1016/j.cpc.2021.108144}
  {\path{doi:10.1016/j.cpc.2021.108144}}.

\bibitem{Lozano:2021zbu}
V.~M. Lozano, R.~M.~S. Seoane, J.~Zurita, {Z'-explorer 2.0: Reconnoitering the
  dark matter landscape}, Comput. Phys. Commun. 288 (2023) 108729.
\newblock \href {http://arxiv.org/abs/2109.13194} {\path{arXiv:2109.13194}},
  \href {https://doi.org/10.1016/j.cpc.2023.108729}
  {\path{doi:10.1016/j.cpc.2023.108729}}.

\bibitem{ATLAS:2018tvr}
{Prospects for searches for heavy $Z^\prime$ and $W^\prime$ bosons in fermionic
  final states with the ATLAS experiment at the HL-LHC} (2018).

\bibitem{Cvetic:1995zs}
M.~Cvetic, S.~Godfrey, {Discovery and identification of extra gauge bosons},
  1995, pp. 383--415.
\newblock \href {http://arxiv.org/abs/hep-ph/9504216}
  {\path{arXiv:hep-ph/9504216}}, \href
  {https://doi.org/10.1142/9789812830265_0007}
  {\path{doi:10.1142/9789812830265_0007}}.

\bibitem{Leike:1996pj}
A.~Leike, S.~Riemann, {$Z^\prime$ search in $e^{+} e^{-}$ annihilation}, Z.
  Phys. C 75 (1997) 341--348.
\newblock \href {http://arxiv.org/abs/hep-ph/9607306}
  {\path{arXiv:hep-ph/9607306}}, \href {https://doi.org/10.1007/s002880050477}
  {\path{doi:10.1007/s002880050477}}.

\bibitem{Godfrey:2005pm}
S.~Godfrey, P.~Kalyniak, A.~Tomkins, {Distinguishing between models with extra
  gauge bosons at the ILC}, in: {2005 International Linear Collider Physics and
  Detector Workshop and 2nd ILC Accelerator Workshop}, 2005.
\newblock \href {http://arxiv.org/abs/hep-ph/0511335}
  {\path{arXiv:hep-ph/0511335}}.

\bibitem{Osland:2009dp}
P.~Osland, A.~A. Pankov, A.~V. Tsytrinov, {Identification of extra neutral
  gauge bosons at the International Linear Collider}, Eur. Phys. J. C 67 (2010)
  191--204.
\newblock \href {http://arxiv.org/abs/0912.2806} {\path{arXiv:0912.2806}},
  \href {https://doi.org/10.1140/epjc/s10052-010-1272-z}
  {\path{doi:10.1140/epjc/s10052-010-1272-z}}.

\bibitem{Andreev:2012cj}
V.~V. Andreev, G.~Moortgat-Pick, P.~Osland, A.~A. Pankov, N.~Paver,
  {Discriminating Z' from Anomalous Trilinear Gauge Coupling Signatures in e+e-
  \textbackslash{}to W+W- at ILC with Polarized Beams}, Eur. Phys. J. C 72
  (2012) 2147.
\newblock \href {http://arxiv.org/abs/1205.0866} {\path{arXiv:1205.0866}},
  \href {https://doi.org/10.1140/epjc/s10052-012-2147-2}
  {\path{doi:10.1140/epjc/s10052-012-2147-2}}.

\bibitem{Han:2013mra}
T.~Han, P.~Langacker, Z.~Liu, L.-T. Wang, {Diagnosis of a New Neutral Gauge
  Boson at the LHC and ILC for Snowmass 2013} (8 2013).
\newblock \href {http://arxiv.org/abs/1308.2738} {\path{arXiv:1308.2738}}.

\bibitem{Godfrey:2013eta}
S.~Godfrey, T.~Martin, {Z' Discovery Reach at Future Hadron Colliders: A
  Snowmass White Paper}, in: {Snowmass 2013}: {Snowmass on the Mississippi},
  2013.
\newblock \href {http://arxiv.org/abs/1309.1688} {\path{arXiv:1309.1688}}.

\bibitem{Kapukchyan:2013hfa}
D.~Kapukchyan, T.~M.~P. Tait, {Sensitivity of a Future High Energy $e^+ e^-$
  Collider to $Z^\prime$ Bosons}, J. Phys. G 41 (2014) 075011.
\newblock \href {http://arxiv.org/abs/1312.3377} {\path{arXiv:1312.3377}},
  \href {https://doi.org/10.1088/0954-3899/41/7/075011}
  {\path{doi:10.1088/0954-3899/41/7/075011}}.

\bibitem{Pankov:2017dkv}
A.~. A.~. Pankov, A.~.~V. Tsytrinov, {Model identification of new heavy
  $Z^\prime$ bosons at ILC with polarized beams}, J. Phys. Conf. Ser. 938~(1)
  (2017) 012059.
\newblock \href {https://doi.org/10.1088/1742-6596/938/1/012059}
  {\path{doi:10.1088/1742-6596/938/1/012059}}.

\bibitem{Gulov:2017uqa}
A.~Gulov, Y.~Moroz, {Optimal observables for $Z'$ models in annihilation
  leptonic processes}, Phys. Rev. D 98~(11) (2018) 115014.
\newblock \href {http://arxiv.org/abs/1711.02853} {\path{arXiv:1711.02853}},
  \href {https://doi.org/10.1103/PhysRevD.98.115014}
  {\path{doi:10.1103/PhysRevD.98.115014}}.

\bibitem{Lu:2023jlr}
Z.~Lu, H.~Li, Z.-L. Han, Z.-G. Si, L.~Zhao, {Phenomenology of Heavy Neutral
  Gauge Boson at Muon Collider} (12 2023).
\newblock \href {http://arxiv.org/abs/2312.17427} {\path{arXiv:2312.17427}}.

\bibitem{MuonCollider:2022glg}
S.~Jindariani, et~al., {Promising Technologies and R\&D Directions for the
  Future Muon Collider Detectors} (3 2022).
\newblock \href {http://arxiv.org/abs/2203.07224} {\path{arXiv:2203.07224}}.

\bibitem{MuonCollider:2022ded}
N.~Bartosik, et~al., {Simulated Detector Performance at the Muon Collider} (3
  2022).
\newblock \href {http://arxiv.org/abs/2203.07964} {\path{arXiv:2203.07964}}.

\bibitem{MuonCollider:2022nsa}
D.~Stratakis, et~al., {A Muon Collider Facility for Physics Discovery} (3
  2022).
\newblock \href {http://arxiv.org/abs/2203.08033} {\path{arXiv:2203.08033}}.

\bibitem{Aime:2022flm}
C.~Aime, et~al., {Muon Collider Physics Summary} (3 2022).
\newblock \href {http://arxiv.org/abs/2203.07256} {\path{arXiv:2203.07256}}.

\bibitem{MuonCollider:2022xlm}
J.~de~Blas, et~al., {The physics case of a 3 TeV muon collider stage} (3 2022).
\newblock \href {http://arxiv.org/abs/2203.07261} {\path{arXiv:2203.07261}}.

\bibitem{P5report}
{2023 P5 Report: Exploring the Quantum Universe},
  \url{https://www.usparticlephysics.org/2023-p5-report/} (2023).

\bibitem{Accettura:2023ked}
C.~Accettura, et~al., {Towards a muon collider}, Eur. Phys. J. C 83~(9) (2023)
  864, [Erratum: Eur.Phys.J.C 84, 36 (2024)].
\newblock \href {http://arxiv.org/abs/2303.08533} {\path{arXiv:2303.08533}},
  \href {https://doi.org/10.1140/epjc/s10052-023-11889-x}
  {\path{doi:10.1140/epjc/s10052-023-11889-x}}.

\bibitem{EuropeanStrategyGroup:2020pow}
{2020 Update of the European Strategy for Particle Physics}, CERN Council,
  Geneva, 2020.
\newblock \href {https://doi.org/10.17181/ESU2020}
  {\path{doi:10.17181/ESU2020}}.

\bibitem{Georgi:1974sy}
H.~Georgi, S.~L. Glashow, {Unity of All Elementary Particle Forces}, Phys. Rev.
  Lett. 32 (1974) 438--441.
\newblock \href {https://doi.org/10.1103/PhysRevLett.32.438}
  {\path{doi:10.1103/PhysRevLett.32.438}}.

\bibitem{Pati:1974yy}
J.~C. Pati, A.~Salam, {Lepton Number as the Fourth Color}, Phys. Rev. D 10
  (1974) 275--289, [Erratum: Phys.Rev.D 11, 703--703 (1975)].
\newblock \href {https://doi.org/10.1103/PhysRevD.10.275}
  {\path{doi:10.1103/PhysRevD.10.275}}.

\bibitem{Gell-Mann:1979vob}
M.~Gell-Mann, P.~Ramond, R.~Slansky, {Complex Spinors and Unified Theories},
  Conf. Proc. C 790927 (1979) 315--321.
\newblock \href {http://arxiv.org/abs/1306.4669} {\path{arXiv:1306.4669}}.

\bibitem{Slansky:1981yr}
R.~Slansky, {Group Theory for Unified Model Building}, Phys. Rept. 79 (1981)
  1--128.
\newblock \href {https://doi.org/10.1016/0370-1573(81)90092-2}
  {\path{doi:10.1016/0370-1573(81)90092-2}}.

\bibitem{Langacker:2000ju}
P.~Langacker, M.~Plumacher, {Flavor changing effects in theories with a heavy
  $Z^\prime$ boson with family nonuniversal couplings}, Phys. Rev. D 62 (2000)
  013006.
\newblock \href {http://arxiv.org/abs/hep-ph/0001204}
  {\path{arXiv:hep-ph/0001204}}, \href
  {https://doi.org/10.1103/PhysRevD.62.013006}
  {\path{doi:10.1103/PhysRevD.62.013006}}.

\bibitem{Baek:2006bv}
S.~Baek, J.~H. Jeon, C.~S. Kim, {B0(s) - anti-B0(s) Mixing in Leptophobic
  Z-prime Model}, Phys. Lett. B 641 (2006) 183--188.
\newblock \href {http://arxiv.org/abs/hep-ph/0607113}
  {\path{arXiv:hep-ph/0607113}}, \href
  {https://doi.org/10.1016/j.physletb.2006.08.041}
  {\path{doi:10.1016/j.physletb.2006.08.041}}.

\bibitem{Ma:1995xk}
E.~Ma, {Neutrino masses in an extended gauge model with E(6) particle content},
  Phys. Lett. B 380 (1996) 286--290.
\newblock \href {http://arxiv.org/abs/hep-ph/9507348}
  {\path{arXiv:hep-ph/9507348}}, \href
  {https://doi.org/10.1016/0370-2693(96)00524-2}
  {\path{doi:10.1016/0370-2693(96)00524-2}}.

\bibitem{Keith:1996fv}
E.~Keith, E.~Ma, {Efficacious extra U(1) factor for the supersymmetric standard
  model}, Phys. Rev. D 54 (1996) 3587--3593.
\newblock \href {http://arxiv.org/abs/hep-ph/9603353}
  {\path{arXiv:hep-ph/9603353}}, \href
  {https://doi.org/10.1103/PhysRevD.54.3587}
  {\path{doi:10.1103/PhysRevD.54.3587}}.

\bibitem{Barger:2003zh}
V.~Barger, P.~Langacker, H.-S. Lee, {Primordial nucleosynthesis constraints on
  $Z^\prime$ properties}, Phys. Rev. D 67 (2003) 075009.
\newblock \href {http://arxiv.org/abs/hep-ph/0302066}
  {\path{arXiv:hep-ph/0302066}}, \href
  {https://doi.org/10.1103/PhysRevD.67.075009}
  {\path{doi:10.1103/PhysRevD.67.075009}}.

\bibitem{Kang:2004ix}
J.-h. Kang, P.~Langacker, T.-j. Li, {Neutrino masses in supersymmetric SU(3)(C)
  x SU(2)(L) x U(1)(Y) x U(1)-prime models}, Phys. Rev. D 71 (2005) 015012.
\newblock \href {http://arxiv.org/abs/hep-ph/0411404}
  {\path{arXiv:hep-ph/0411404}}, \href
  {https://doi.org/10.1103/PhysRevD.71.015012}
  {\path{doi:10.1103/PhysRevD.71.015012}}.

\bibitem{King:2005jy}
S.~F. King, S.~Moretti, R.~Nevzorov, {Theory and phenomenology of an
  exceptional supersymmetric standard model}, Phys. Rev. D 73 (2006) 035009.
\newblock \href {http://arxiv.org/abs/hep-ph/0510419}
  {\path{arXiv:hep-ph/0510419}}, \href
  {https://doi.org/10.1103/PhysRevD.73.035009}
  {\path{doi:10.1103/PhysRevD.73.035009}}.

\bibitem{Mohapatra:1974gc}
R.~N. Mohapatra, J.~C. Pati, {A Natural Left-Right Symmetry}, Phys. Rev. D 11
  (1975) 2558.
\newblock \href {https://doi.org/10.1103/PhysRevD.11.2558}
  {\path{doi:10.1103/PhysRevD.11.2558}}.

\bibitem{deCarlos:1997yv}
B.~de~Carlos, J.~R. Espinosa, {Cold dark matter candidate in a class of
  supersymmetric models with an extra U(1)}, Phys. Lett. B 407 (1997) 12--21.
\newblock \href {http://arxiv.org/abs/hep-ph/9705315}
  {\path{arXiv:hep-ph/9705315}}, \href
  {https://doi.org/10.1016/S0370-2693(97)00747-8}
  {\path{doi:10.1016/S0370-2693(97)00747-8}}.

\bibitem{Nakamura:2006ht}
S.~Nakamura, D.~Suematsu, {Supersymmetric extra U(1) models with a singlino
  dominated LSP}, Phys. Rev. D 75 (2007) 055004.
\newblock \href {http://arxiv.org/abs/hep-ph/0609061}
  {\path{arXiv:hep-ph/0609061}}, \href
  {https://doi.org/10.1103/PhysRevD.75.055004}
  {\path{doi:10.1103/PhysRevD.75.055004}}.

\bibitem{Barger:2007nv}
V.~Barger, P.~Langacker, I.~Lewis, M.~McCaskey, G.~Shaughnessy, B.~Yencho,
  {Recoil Detection of the Lightest Neutralino in MSSM Singlet Extensions},
  Phys. Rev. D 75 (2007) 115002.
\newblock \href {http://arxiv.org/abs/hep-ph/0702036}
  {\path{arXiv:hep-ph/0702036}}, \href
  {https://doi.org/10.1103/PhysRevD.75.115002}
  {\path{doi:10.1103/PhysRevD.75.115002}}.

\bibitem{Lee:2007mt}
H.-S. Lee, K.~T. Matchev, S.~Nasri, {Revival of the thermal sneutrino dark
  matter}, Phys. Rev. D 76 (2007) 041302.
\newblock \href {http://arxiv.org/abs/hep-ph/0702223}
  {\path{arXiv:hep-ph/0702223}}, \href
  {https://doi.org/10.1103/PhysRevD.76.041302}
  {\path{doi:10.1103/PhysRevD.76.041302}}.

\bibitem{Belanger:2007dx}
G.~Belanger, A.~Pukhov, G.~Servant, {Dirac Neutrino Dark Matter}, JCAP 01
  (2008) 009.
\newblock \href {http://arxiv.org/abs/0706.0526} {\path{arXiv:0706.0526}},
  \href {https://doi.org/10.1088/1475-7516/2008/01/009}
  {\path{doi:10.1088/1475-7516/2008/01/009}}.

\bibitem{Hewett:1988xc}
J.~L. Hewett, T.~G. Rizzo, {Low-Energy Phenomenology of Superstring Inspired
  E(6) Models}, Phys. Rept. 183 (1989) 193.
\newblock \href {https://doi.org/10.1016/0370-1573(89)90071-9}
  {\path{doi:10.1016/0370-1573(89)90071-9}}.

\bibitem{Barger:1980dx}
V.~D. Barger, W.-Y. Keung, E.~Ma, {Sequential $W$ and $Z$ Bosons}, Phys. Lett.
  B 94 (1980) 377--380.
\newblock \href {https://doi.org/10.1016/0370-2693(80)90900-4}
  {\path{doi:10.1016/0370-2693(80)90900-4}}.

\bibitem{Robinett:1981yz}
R.~W. Robinett, J.~L. Rosner, {Prospects for a Second Neutral Vector Boson at
  Low Mass in SO(10)}, Phys. Rev. D 25 (1982) 3036, [Erratum: Phys.Rev.D 27,
  679 (1983)].
\newblock \href {https://doi.org/10.1103/PhysRevD.27.679}
  {\path{doi:10.1103/PhysRevD.27.679}}.

\bibitem{Altarelli:1989ff}
G.~Altarelli, B.~Mele, M.~Ruiz-Altaba, {Searching for New Heavy Vector Bosons
  in $p \bar{p}$ Colliders}, Z. Phys. C 45 (1989) 109, [Erratum: Z.Phys.C 47,
  676 (1990)].
\newblock \href {https://doi.org/10.1007/BF01556677}
  {\path{doi:10.1007/BF01556677}}.

\bibitem{achiman1978quark}
Y.~Achiman, B.~Stech, Quark-lepton symmetry and mass scales in an e6 unified
  gauge model, Physics Letters B 77~(4-5) (1978) 389--393.

\bibitem{London:1986dk}
D.~London, J.~L. Rosner, {Extra Gauge Bosons in E(6)}, Phys. Rev. D 34 (1986)
  1530.
\newblock \href {https://doi.org/10.1103/PhysRevD.34.1530}
  {\path{doi:10.1103/PhysRevD.34.1530}}.

\bibitem{Mohapatra:1979ia}
R.~N. Mohapatra, G.~Senjanovic, {Neutrino Mass and Spontaneous Parity
  Nonconservation}, Phys. Rev. Lett. 44 (1980) 912.
\newblock \href {https://doi.org/10.1103/PhysRevLett.44.912}
  {\path{doi:10.1103/PhysRevLett.44.912}}.

\bibitem{Mohapatra:1980yp}
R.~N. Mohapatra, G.~Senjanovic, {Neutrino Masses and Mixings in Gauge Models
  with Spontaneous Parity Violation}, Phys. Rev. D 23 (1981) 165.
\newblock \href {https://doi.org/10.1103/PhysRevD.23.165}
  {\path{doi:10.1103/PhysRevD.23.165}}.

\bibitem{PhysRevD.36.274}
E.~Ma, \href{https://link.aps.org/doi/10.1103/PhysRevD.36.274}{Particle
  dichotomy and left-right decomposition of ${\mathrm{e}}_{6}$ superstring
  models}, Phys. Rev. D 36 (1987) 274--276.
\newblock \href {https://doi.org/10.1103/PhysRevD.36.274}
  {\path{doi:10.1103/PhysRevD.36.274}}.
\newline\urlprefix\url{https://link.aps.org/doi/10.1103/PhysRevD.36.274}

\bibitem{Ashry_2015}
M.~Ashry, S.~Khalil,
  \href{https://doi.org/10.1103%2Fphysrevd.91.015009}{Phenomenological aspects
  of a {TeV}-scale alternative left-right model}, Physical Review D 91~(1) (jan
  2015).
\newblock \href {https://doi.org/10.1103/physrevd.91.015009}
  {\path{doi:10.1103/physrevd.91.015009}}.
\newline\urlprefix\url{https://doi.org/10.1103%2Fphysrevd.91.015009}

\bibitem{Arkani_Hamed_2002}
N.~Arkani-Hamed, A.~G. Cohen, E.~Katz, A.~E. Nelson, {The Littlest Higgs}, JHEP
  07 (2002) 034.
\newblock \href {http://arxiv.org/abs/hep-ph/0206021}
  {\path{arXiv:hep-ph/0206021}}, \href
  {https://doi.org/10.1088/1126-6708/2002/07/034}
  {\path{doi:10.1088/1126-6708/2002/07/034}}.

\bibitem{Han_2003}
T.~Han, H.~E. Logan, B.~McElrath, L.-T. Wang, {Phenomenology of the little
  Higgs model}, Phys. Rev. D 67 (2003) 095004.
\newblock \href {http://arxiv.org/abs/hep-ph/0301040}
  {\path{arXiv:hep-ph/0301040}}, \href
  {https://doi.org/10.1103/PhysRevD.67.095004}
  {\path{doi:10.1103/PhysRevD.67.095004}}.

\bibitem{Kilian:2003xt}
W.~Kilian, J.~Reuter, {The Low-energy structure of little Higgs models}, Phys.
  Rev. D 70 (2004) 015004.
\newblock \href {http://arxiv.org/abs/hep-ph/0311095}
  {\path{arXiv:hep-ph/0311095}}, \href
  {https://doi.org/10.1103/PhysRevD.70.015004}
  {\path{doi:10.1103/PhysRevD.70.015004}}.

\bibitem{Schmaltz_2004}
M.~Schmaltz, {The Simplest little Higgs}, JHEP 08 (2004) 056.
\newblock \href {http://arxiv.org/abs/hep-ph/0407143}
  {\path{arXiv:hep-ph/0407143}}, \href
  {https://doi.org/10.1088/1126-6708/2004/08/056}
  {\path{doi:10.1088/1126-6708/2004/08/056}}.

\bibitem{Kilian:2004pp}
W.~Kilian, D.~Rainwater, J.~Reuter, {Pseudo-axions in little Higgs models},
  Phys. Rev. D 71 (2005) 015008.
\newblock \href {http://arxiv.org/abs/hep-ph/0411213}
  {\path{arXiv:hep-ph/0411213}}, \href
  {https://doi.org/10.1103/PhysRevD.71.015008}
  {\path{doi:10.1103/PhysRevD.71.015008}}.

\bibitem{Kilian:2006eh}
W.~Kilian, D.~Rainwater, J.~Reuter, {Distinguishing little-Higgs product and
  simple group models at the LHC and ILC}, Phys. Rev. D 74 (2006) 095003,
  [Erratum: Phys.Rev.D 74, 099905 (2006)].
\newblock \href {http://arxiv.org/abs/hep-ph/0609119}
  {\path{arXiv:hep-ph/0609119}}, \href
  {https://doi.org/10.1103/PhysRevD.74.095003}
  {\path{doi:10.1103/PhysRevD.74.095003}}.

\bibitem{Oda:2015gna}
S.~Oda, N.~Okada, D.-s. Takahashi, {Classically conformal U(1)' extended
  standard model and Higgs vacuum stability}, Phys. Rev. D 92~(1) (2015)
  015026.
\newblock \href {http://arxiv.org/abs/1504.06291} {\path{arXiv:1504.06291}},
  \href {https://doi.org/10.1103/PhysRevD.92.015026}
  {\path{doi:10.1103/PhysRevD.92.015026}}.

\bibitem{iso2009classically}
S.~Iso, N.~Okada, Y.~Orikasa, Classically conformal b--l extended standard
  model, Physics Letters B 676~(1-3) (2009) 81--87.

\bibitem{Fritzsch:1974nn}
H.~Fritzsch, P.~Minkowski, {Unified Interactions of Leptons and Hadrons},
  Annals Phys. 93 (1975) 193--266.
\newblock \href {https://doi.org/10.1016/0003-4916(75)90211-0}
  {\path{doi:10.1016/0003-4916(75)90211-0}}.

\bibitem{Glashow:1984gc}
A.~de~R\'ujula, H.~Georgi, S.~L. Glashow, {Trinification of All Elementary
  Particle Forces}, in: {Fifth Workshop on Grand Unification}, 1984.

\bibitem{Kilian:2006hh}
W.~Kilian, J.~Reuter, {Unification without doublet-triplet splitting}, Phys.
  Lett. B 642 (2006) 81--84.
\newblock \href {http://arxiv.org/abs/hep-ph/0606277}
  {\path{arXiv:hep-ph/0606277}}, \href
  {https://doi.org/10.1016/j.physletb.2006.09.033}
  {\path{doi:10.1016/j.physletb.2006.09.033}}.

\bibitem{Braam:2010sy}
F.~Braam, A.~Knochel, J.~Reuter, {An Exceptional SSM from E6 Orbifold GUTs with
  intermediate LR symmetry}, JHEP 06 (2010) 013.
\newblock \href {http://arxiv.org/abs/1001.4074} {\path{arXiv:1001.4074}},
  \href {https://doi.org/10.1007/JHEP06(2010)013}
  {\path{doi:10.1007/JHEP06(2010)013}}.

\bibitem{Braam:2011xh}
F.~Braam, J.~Reuter, {A Simplified Scheme for GUT-inspired Theories with
  Multiple Abelian Factors}, Eur. Phys. J. C 72 (2012) 1885.
\newblock \href {http://arxiv.org/abs/1107.2806} {\path{arXiv:1107.2806}},
  \href {https://doi.org/10.1140/epjc/s10052-012-1885-5}
  {\path{doi:10.1140/epjc/s10052-012-1885-5}}.

\bibitem{Rizzo:2012rf}
T.~G. Rizzo, {Gauge Kinetic Mixing in the $E_6$SSM}, Phys. Rev. D 85 (2012)
  055010.
\newblock \href {http://arxiv.org/abs/1201.2898} {\path{arXiv:1201.2898}},
  \href {https://doi.org/10.1103/PhysRevD.85.055010}
  {\path{doi:10.1103/PhysRevD.85.055010}}.

\bibitem{Randall:1999ee}
L.~Randall, R.~Sundrum, {A Large mass hierarchy from a small extra dimension},
  Phys. Rev. Lett. 83 (1999) 3370--3373.
\newblock \href {http://arxiv.org/abs/hep-ph/9905221}
  {\path{arXiv:hep-ph/9905221}}, \href
  {https://doi.org/10.1103/PhysRevLett.83.3370}
  {\path{doi:10.1103/PhysRevLett.83.3370}}.

\bibitem{Appelquist:2000nn}
T.~Appelquist, H.-C. Cheng, B.~A. Dobrescu, {Bounds on universal extra
  dimensions}, Phys. Rev. D 64 (2001) 035002.
\newblock \href {http://arxiv.org/abs/hep-ph/0012100}
  {\path{arXiv:hep-ph/0012100}}, \href
  {https://doi.org/10.1103/PhysRevD.64.035002}
  {\path{doi:10.1103/PhysRevD.64.035002}}.

\bibitem{Han:1998sg}
T.~Han, J.~D. Lykken, R.-J. Zhang, {On Kaluza-Klein states from large extra
  dimensions}, Phys. Rev. D 59 (1999) 105006.
\newblock \href {http://arxiv.org/abs/hep-ph/9811350}
  {\path{arXiv:hep-ph/9811350}}, \href
  {https://doi.org/10.1103/PhysRevD.59.105006}
  {\path{doi:10.1103/PhysRevD.59.105006}}.

\bibitem{ALEPH:2001uca}
A.~Heister, et~al., {Measurement of the tau polarization at LEP}, Eur. Phys. J.
  C 20 (2001) 401--430.
\newblock \href {http://arxiv.org/abs/hep-ex/0104038}
  {\path{arXiv:hep-ex/0104038}}, \href {https://doi.org/10.1007/s100520100689}
  {\path{doi:10.1007/s100520100689}}.

\bibitem{DELPHI:1999yne}
P.~Abreu, et~al., {A Precise measurement of the tau polarization at LEP-1},
  Eur. Phys. J. C 14 (2000) 585--611.
\newblock \href {https://doi.org/10.1007/s100520000363}
  {\path{doi:10.1007/s100520000363}}.

\bibitem{talk_Epiphany}
K.~Mekala, D.~Jeans, J.~Tian, J.~Reuter, A.~F. Żarnecki, Precise measurement
  of light-quark electroweak couplings at future colliders, presented at the
  Epiphany'24 conference,
  \url{https://indico.cern.ch/event/1288528/contributions/5718576/} (2024).

\bibitem{ZurbanoFernandez:2020cco}
I.~Zurbano~Fernandez, et~al., {High-Luminosity Large Hadron Collider (HL-LHC):
  Technical design report} 10/2020 (12 2020).
\newblock \href {https://doi.org/10.23731/CYRM-2020-0010}
  {\path{doi:10.23731/CYRM-2020-0010}}.

\bibitem{ILC:2013jhg}
{The International Linear Collider Technical Design Report - Volume 2: Physics}
  (6 2013).
\newblock \href {http://arxiv.org/abs/1306.6352} {\path{arXiv:1306.6352}}.

\bibitem{Behnke:2013lya}
H.~Abramowicz, et~al., {The International Linear Collider Technical Design
  Report - Volume 4: Detectors} (6 2013).
\newblock \href {http://arxiv.org/abs/1306.6329} {\path{arXiv:1306.6329}}.

\bibitem{FCC:2018evy}
A.~Abada, et~al., {FCC-ee: The Lepton Collider}: {Future Circular Collider
  Conceptual Design Report Volume 2}, Eur. Phys. J. ST 228~(2) (2019) 261--623.
\newblock \href {https://doi.org/10.1140/epjst/e2019-900045-4}
  {\path{doi:10.1140/epjst/e2019-900045-4}}.

\bibitem{CEPCStudyGroup:2023quu}
W.~Abdallah, et~al., {CEPC Technical Design Report -- Accelerator (v2)} (12
  2023).
\newblock \href {http://arxiv.org/abs/2312.14363} {\path{arXiv:2312.14363}}.

\bibitem{Aicheler:2012bya}
{A Multi-TeV Linear Collider Based on CLIC Technology}: {CLIC Conceptual Design
  Report} (10 2012).
\newblock \href {https://doi.org/10.5170/CERN-2012-007}
  {\path{doi:10.5170/CERN-2012-007}}.

\bibitem{Linssen:2012hp}
{Physics and Detectors at CLIC: CLIC Conceptual Design Report} (2 2012).
\newblock \href {http://arxiv.org/abs/1202.5940} {\path{arXiv:1202.5940}},
  \href {https://doi.org/10.5170/CERN-2012-003}
  {\path{doi:10.5170/CERN-2012-003}}.

\bibitem{FCC:2018vvp}
A.~Abada, et~al., {FCC-hh: The Hadron Collider}: {Future Circular Collider
  Conceptual Design Report Volume 3}, Eur. Phys. J. ST 228~(4) (2019)
  755--1107.
\newblock \href {https://doi.org/10.1140/epjst/e2019-900087-0}
  {\path{doi:10.1140/epjst/e2019-900087-0}}.

\bibitem{FCC:2018byv}
A.~Abada, et~al., {FCC Physics Opportunities}: {Future Circular Collider
  Conceptual Design Report Volume 1}, Eur. Phys. J. C 79~(6) (2019) 474.
\newblock \href {https://doi.org/10.1140/epjc/s10052-019-6904-3}
  {\path{doi:10.1140/epjc/s10052-019-6904-3}}.

\bibitem{Tang:2015qga}
J.~Tang, et~al., {Concept for a Future Super Proton-Proton Collider} (7 2015).
\newblock \href {http://arxiv.org/abs/1507.03224} {\path{arXiv:1507.03224}}.

\bibitem{MICE:2019jkl}
M.~Bogomilov, et~al., {Demonstration of cooling by the Muon Ionization Cooling
  Experiment}, Nature 578~(7793) (2020) 53--59.
\newblock \href {http://arxiv.org/abs/1907.08562} {\path{arXiv:1907.08562}},
  \href {https://doi.org/10.1038/s41586-020-1958-9}
  {\path{doi:10.1038/s41586-020-1958-9}}.

\bibitem{Chen:2016wkt}
J.~Chen, T.~Han, B.~Tweedie, {Electroweak Splitting Functions and High Energy
  Showering}, JHEP 11 (2017) 093.
\newblock \href {http://arxiv.org/abs/1611.00788} {\path{arXiv:1611.00788}},
  \href {https://doi.org/10.1007/JHEP11(2017)093}
  {\path{doi:10.1007/JHEP11(2017)093}}.

\bibitem{Han:2020uid}
T.~Han, Y.~Ma, K.~Xie, {High energy leptonic collisions and electroweak parton
  distribution functions}, Phys. Rev. D 103~(3) (2021) L031301.
\newblock \href {http://arxiv.org/abs/2007.14300} {\path{arXiv:2007.14300}},
  \href {https://doi.org/10.1103/PhysRevD.103.L031301}
  {\path{doi:10.1103/PhysRevD.103.L031301}}.

\bibitem{Garosi:2023bvq}
F.~Garosi, D.~Marzocca, S.~Trifinopoulos, {LePDF: Standard Model PDFs for
  high-energy lepton colliders}, JHEP 09 (2023) 107.
\newblock \href {http://arxiv.org/abs/2303.16964} {\path{arXiv:2303.16964}},
  \href {https://doi.org/10.1007/JHEP09(2023)107}
  {\path{doi:10.1007/JHEP09(2023)107}}.

\bibitem{Kilian_2011}
W.~Kilian, T.~Ohl, J.~Reuter,
  \href{http://dx.doi.org/10.1140/epjc/s10052-011-1742-y}{Whizard—simulating
  multi-particle processes at lhc and ilc}, The European Physical Journal C
  71~(9) (Sep. 2011).
\newblock \href {https://doi.org/10.1140/epjc/s10052-011-1742-y}
  {\path{doi:10.1140/epjc/s10052-011-1742-y}}.
\newline\urlprefix\url{http://dx.doi.org/10.1140/epjc/s10052-011-1742-y}

\bibitem{moretti2001omega}
M.~Moretti, T.~Ohl, J.~Reuter, O'mega: An optimizing matrix element generator
  (2001).
\newblock \href {http://arxiv.org/abs/hep-ph/0102195}
  {\path{arXiv:hep-ph/0102195}}.

\bibitem{Christensen:2010wz}
N.~D. Christensen, C.~Duhr, B.~Fuks, J.~Reuter, C.~Speckner, {Introducing an
  interface between WHIZARD and FeynRules}, Eur. Phys. J. C 72 (2012) 1990.
\newblock \href {http://arxiv.org/abs/1010.3251} {\path{arXiv:1010.3251}},
  \href {https://doi.org/10.1140/epjc/s10052-012-1990-5}
  {\path{doi:10.1140/epjc/s10052-012-1990-5}}.

\bibitem{Degrande:2011ua}
C.~Degrande, C.~Duhr, B.~Fuks, D.~Grellscheid, O.~Mattelaer, T.~Reiter, {UFO -
  The Universal FeynRules Output}, Comput. Phys. Commun. 183 (2012) 1201--1214.
\newblock \href {http://arxiv.org/abs/1108.2040} {\path{arXiv:1108.2040}},
  \href {https://doi.org/10.1016/j.cpc.2012.01.022}
  {\path{doi:10.1016/j.cpc.2012.01.022}}.

\bibitem{Darme:2023jdn}
L.~Darm\'e, et~al., {UFO 2.0: the \textquoteleft{}Universal Feynman
  Output\textquoteright{} format}, Eur. Phys. J. C 83~(7) (2023) 631.
\newblock \href {http://arxiv.org/abs/2304.09883} {\path{arXiv:2304.09883}},
  \href {https://doi.org/10.1140/epjc/s10052-023-11780-9}
  {\path{doi:10.1140/epjc/s10052-023-11780-9}}.

\bibitem{Bredt:2022dmm}
P.~M. Bredt, W.~Kilian, J.~Reuter, P.~Stienemeier, {NLO electroweak corrections
  to multi-boson processes at a muon collider}, JHEP 12 (2022) 138.
\newblock \href {http://arxiv.org/abs/2208.09438} {\path{arXiv:2208.09438}},
  \href {https://doi.org/10.1007/JHEP12(2022)138}
  {\path{doi:10.1007/JHEP12(2022)138}}.

\end{thebibliography}

\end{document}